\documentclass[aps,prx,twocolumn,superscriptaddress,longbibliography]{revtex4-1}

\usepackage{amsmath}
\usepackage{amssymb}
\usepackage{graphicx}
\usepackage{afterpage}
\usepackage{amsfonts}
\usepackage{amssymb}
\usepackage{graphicx}
\usepackage{braket}
\usepackage{textcomp}

\newcommand{\be}{\begin{equation}}
\newcommand{\ee}{\end{equation}}
\newcommand{\bea}{\begin{eqnarray}}
\newcommand{\eea}{\end{eqnarray}}

\newcommand{\polho}{\relbar}
\newcommand{\polve}{|}
\newcommand{\polcp}{\circlearrowleft} 
\newcommand{\polcm}{\circlearrowright} 
\newcommand{\poldu}{\diagup}\newcommand{\poldd}{\diagdown}

\newcommand{\spl}{\sigma^{+}}
\newcommand{\smi}{\sigma^{-}}
\newcommand{\spm}{\sigma^{\pm}}

\newcommand{\proj}[2]{\left|#1\right>\left<#2\right|}

\begin{document}

\title{Dynamics of excitons in individual InAs~quantum dots\\ revealed in four-wave mixing spectroscopy}





\author{Q.~Mermillod}
\email{quentin.mermillod@neel.cnrs.fr}
\affiliation{Univ. Grenoble Alpes, F-38000 Grenoble, France}
\affiliation{CNRS, Institut N\'{e}el, "Nanophysique et semiconducteurs" group, F-38000 Grenoble, France}

\author{D.~Wigger}
\email{d.wigger@wwu.de}
\affiliation{Institut f\"{u}r Festk\"{o}rpertheorie, Universit\"{a}t M\"{u}nster, 48149 M\"{u}nster, Germany}

\author{V.~Delmonte}
\affiliation{Univ. Grenoble Alpes, F-38000 Grenoble, France}
\affiliation{CNRS, Institut N\'{e}el, "Nanophysique et semiconducteurs" group, F-38000 Grenoble, France}

\author{D.~E.~Reiter}
\email{doris.reiter@wwu.de}
\affiliation{Institut f\"{u}r Festk\"{o}rpertheorie, Universit\"{a}t M\"{u}nster, 48149 M\"{u}nster, Germany}

\author{C.~Schneider}
\author{M.~Kamp}
\affiliation{Technische Physik and Wilhelm Conrad R\"{o}ntgen
Research Center for Complex Material Systems, Universit\"{a}t
W\"{u}rzburg, Germany}

\author{S.~H\"{o}fling}
\affiliation{Technische Physik and Wilhelm Conrad R\"{o}ntgen
Research Center for Complex Material Systems, Universit\"{a}t
W\"{u}rzburg, Germany} \affiliation{SUPA, School of Physics and
Astronomy, University of St Andrews, St Andrews, KY16 9SS, United
Kingdom}

\author{W.~Langbein}
\affiliation{Cardiff University School of Physics and Astronomy, The Parade, Cardiff CF24 3AA, United Kingdom}

\author{T.~Kuhn}
\affiliation{Institut f\"{u}r Festk\"{o}rpertheorie, Universit\"{a}t M\"{u}nster, 48149 M\"{u}nster, Germany}

\author{G.~Nogues}
\affiliation{Univ. Grenoble Alpes, F-38000 Grenoble, France}
\affiliation{CNRS, Institut N\'{e}el, "Nanophysique et semiconducteurs" group, F-38000 Grenoble, France}

\author{J.~Kasprzak}
\email[]{jacek.kasprzak@neel.cnrs.fr}
\affiliation{Univ. Grenoble Alpes, F-38000 Grenoble, France}
\affiliation{CNRS, Institut N\'{e}el, "Nanophysique et semiconducteurs" group, F-38000 Grenoble, France}

\begin{abstract}
A detailed understanding of the population and coherence dynamics in
optically driven individual emitters in solids and their signatures
in ultrafast nonlinear-optical signals is of prime importance for
their applications in future quantum and optical technologies. In a
combined experimental and theoretical study on exciton complexes in
single semiconductor quantum dots we reveal a detailed picture of
the dynamics employing three-beam polarization-resolved four-wave
mixing (FWM) micro-spectroscopy. The oscillatory dynamics of the FWM
signals in the exciton-biexciton system is governed by the
fine-structure splitting and the biexciton binding energy in an
excellent quantitative agreement between measurement and analytical
description. The analysis of the excitation conditions exhibits a
dependence of the dynamics on the specific choice of polarization
configuration, pulse areas and temporal ordering of driving fields.
The interplay between the transitions in the four-level exciton
system leads to rich evolution of coherence and population. Using
two-dimensional FWM spectroscopy we elucidate the exciton-biexciton
coupling and identify neutral and charged exciton complexes in a
single quantum dot. Our investigations thus clearly reveal that FWM
spectroscopy is a powerful tool to characterize spectral and
dynamical properties of single quantum structures.
\end{abstract}


\maketitle

\section{Introduction}

\begin{figure}[htbp]
\centering
\includegraphics[width=1\columnwidth]{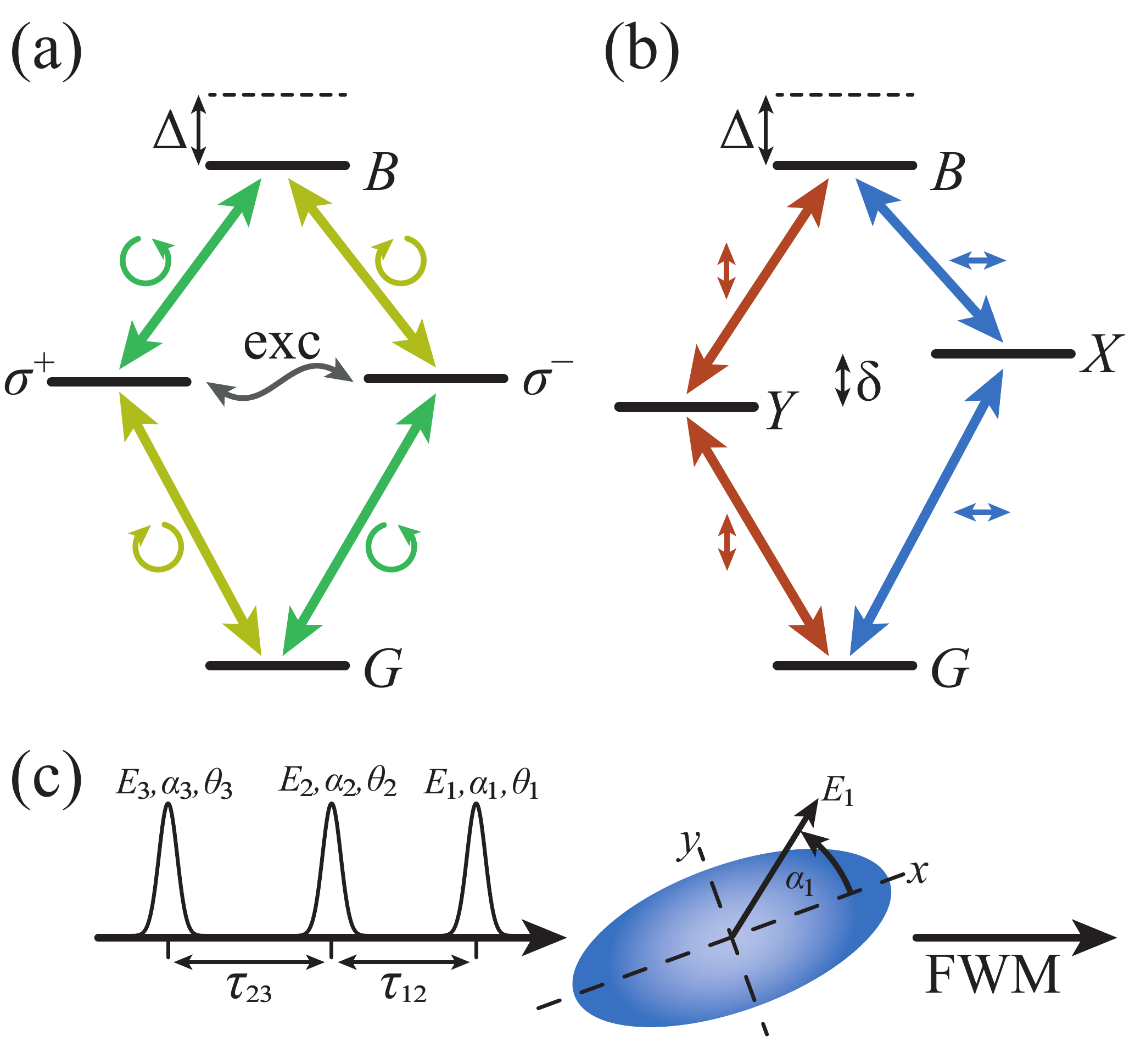}
\caption{{\bf Schematic picture.} Sketches of the QD system for
({\bf a}) circularly polarized excitation and ({\bf b}) linearly
polarized excitation. ({\bf c}) Cartoon of the FWM experiment and
the QD orientation. \label{fig:schema}}
\end{figure}

A comprehensive understanding of exciton complexes and their
transitions in semiconductor quantum dots (QDs) is a crucial step
for assessing their functionality as optically controllable solid
state devices in quantum information technology~\cite{BonadeoSCI98,
Michler03, RamsaySST10}. For example, for ultrafast manipulation of
QDs it is essential to know the decay and decoherence times of
exciton and biexciton~\cite{MonnielloPRL13, AkimovPRL06,
BacherPRL99}, while for entangled photon generation, the fine
structure splitting (FSS) between the excitons and the biexciton
binding energy (BBE) play a decisive role~\cite{SeguinPRL05,
YoungPRB05, StevensonNAT06, AkopianPRL06,DingPRL10}. Also the
alignment between the polarization of the light and the QD axis
becomes important when selectively addressing different
transitions~\cite{ToninPRB12}. In our study we can access all these
quantities, i.e., dephasing and lifetimes, FSS, BBE and dot axis of
a single QD, within the same set-up. For the experimental
investigations we implement a heterodyne spectral interferometry
technique~\cite{LangbeinOL06} to retrieve polarization-resolved
four-wave mixing (FWM) signals. While FWM has often been used to
infer exciton dynamics in QWs~\cite{VossPRB06} or for QD
ensembles~\cite{MoodyPRB13}, for single QDs the experiments are more
challenging because of the weak signal intensity. This long-standing
issue has recently been solved by exploiting photonic nanostructures
to enhance non-linear responses. Employing low-Q planar
microcavities~\cite{Fras15}, conical photonic waveguide
antennas~\cite{Mermillod16} and deterministic
micro-lenses~\cite{GschreyNatComm15} the detection sensitivity of
FWM generated by an exciton is improved by up to four orders of
magnitude with respect to QDs in bulk material. To model the data we
use a density matrix formalism including all populations and
coherences, as well as decay and decoherence
rates~\cite{VagovPRB02,AxtPRB05}. In this paper we present a
comprehensive set of measurements and simulations, exploring the
oscillatory dynamics of coherences and populations in the
exciton-biexciton system in a single QD.

Because the QDs in our sample exhibit charge fluctuation, we can
also study coherent dynamics of more involved complexes, namely
charged excitons. To discriminate between neutral and charged
excitons we use two-dimensional (2D) FWM spectroscopy. In 2D FWM
spectra transitions correspond to peaks on the diagonal, while the
coupling between different states can be seen by off-diagonal peaks
connecting the diagonal
ones~\cite{KasprzakNPho11,CundiffJOSAB12,DaiPRL12}. We show that
this technique allows for a fast, comprehensive characterization of
exciton complexes.

The paper is organized as follows: First, we focus on the
exciton-biexciton system introducing the states, their energies and
couplings. After characterizing quantum beats induced by the FSS and
the BBE, we analyze the angle dependence of the dynamics. In a next
step, we look at the population dynamics revealing coherences that
are typically hidden. Finally, we study 2D FWM maps identifying
neutral and charged exciton complexes.

\section{Exciton-Biexciton System}
\subsection{QD system and FWM}

In this work, we perform FWM measurements of strongly-confined
excitons in individual InAs QDs, embedded in a low-Q asymmetric
GaAs/AlGaAs micro-cavity. A detailed description of the sample
growth and characterization can be found in the Methods section. In
the first part we restrict ourselves to the neutral s-shell excitons
in the QD. Depending on the polarization of the exciting pulses,
excitons with different polarizations are created. For circular
polarization denoted by $\polcm$ and $\polcp$, the system is
characterized by the ground state $|G\rangle$, the two exciton
states $|\sigma^+\rangle$ and $|\sigma^-\rangle$ and the biexciton
state $|B\rangle$ as depicted in Fig.~\ref{fig:schema}\,a. The
energy of the biexciton state is reduced with respect to the double
exciton energy by the BBE, denoted as $\Delta$. The degeneracy of
the two excitons is lifted by the anisotropic confinement potential
of the QD and its zinc-blende crystal structure leading to an
exchange coupling between the excitons. Therefore, the circularly
polarized states are not the energy eigenstates of the system.
Instead, the eigenstates are given by the linearly polarized
excitons $|X\rangle =  \left( |\spl\rangle + |\smi\rangle
\right)/\sqrt{2}$ and $|Y\rangle= i \left( |\spl\rangle -
|\smi\rangle \right)/\sqrt{2}$, which are split by the FSS, labeled
as $\delta$ (see Fig.~\ref{fig:schema}\,b). We define the
polarization axis of the $X$-exciton as the $x$-axis of the QD and
specify the angle of a linearly polarized excitation with respect to
this axis, i.e., for $\alpha=0^\circ$ the $X$-exciton is excited,
while for $\alpha=90^\circ$, the $Y$-exciton is generated as
depicted in Fig.~\ref{fig:schema}\,c. In these particular cases, the
four-level system of the QD can be restricted to three-levels. For
any intermediate angle $\alpha$ a linear combination of $X$- and
$Y$-exciton is created. We will indicate the light polarization
angles $\alpha=(0^{\circ},\,45^{\circ},\,90^{\circ},\,135^{\circ})$
by $(\polho,\,\poldu,\,\polve,\,\poldd)$.

In practice, three laser pulses $E_i$ ($i=1,2,3$) with pulse areas
$\theta_i$ and polarization angles $\alpha_i$ drive FWM of the QD,
as depicted in Fig.~\ref{fig:schema}\,c. The optical frequencies of
$E_i$ are shifted by $\Omega_{i}$ using acousto-optic modulation in
the radio-frequency range. The polarizations of the beams are
adjusted by a set of $\lambda/2$ and $\lambda/4$ plates. The delays
between the pulses are denoted by $\tau_{12}$ and $\tau_{23}$, which
are handled by a pair of mechanical delay stages. A positive delay
corresponds to the case when the first pulse arrives before the
second and so on. The beams are then recombined into the same
spatial mode and focused onto the top of the sample placed in a
cryostat operating at $T=5$~K using an external microscope objective
optimized for NIR spectral range (NA$=0.65$), installed on a
XYZ-piezo stage. The investigated nonlinear polarization is
retrieved by detecting the corresponding phase modulation in
reflectance using a heterodyne technique with a reference beam
$E_{r}$. Our detection scheme combines optical heterodyning with
spectral interferometry, as detailed in Ref.~\cite{LangbeinRNC10}.
The current implementation is presented in Ref.~\cite{Fras15}.

With FWM, we can probe population and coherence dynamics of a QD
exciton. To detect the latter, the time delay $\tau_{23}$ between
$E_2$ and $E_3$ is set to zero, i.e., we only use $E_1$ and $E_2$
and heterodyne at $2\Omega_2-\Omega_1$: the first arriving pulse
creates a coherence in the system, which after $\tau_{12}$ is
transformed into FWM signals by the second pulse. The ratio of their
pulse areas is taken to be $\theta_2=2\theta_1$. To explore the
population dynamics, an excitation with three beams is required and
we look at the heterodyne signal at $(\Omega_3+\Omega_2-\Omega_1)$.
All pulses have the same area $\theta_i=\theta$. The first laser
pulse creates the coherence. The second one, which follows shortly
after at $\tau_{12}=0.5$~ps (yet beyond the overlap of $E_1$ and
$E_2$ to avoid the generation of non-resonant nonlinearities),
creates populations evolving during $\tau_{23}$. The FWM signals are
then launched by the third pulse, $E_3$.

In the calculations the density matrix elements are denoted by
$\varrho_{\nu \nu'}$ with $\nu \in \{ G,\spl,\smi,B \}$ in the
circularly polarized basis and $ \nu \in \{ G,X,Y,B \}$ in the
linearly polarized basis. The equations of motion under an
excitation with a series of ultrafast pulses are solved analytically
following Refs.~\cite{VagovPRB02} and \cite{AxtPRB05}. In between
the pulses, the coherences and populations are subject to decay and
decoherence, where the following rates are taken into account: The
decay of the populations is included by the rate $\gamma=1/T_1$ for
all transitions between biexciton and exciton, as well as between
exciton and ground state, where $T_1$ is the radiative lifetime. The
corresponding coherences $\varrho_{GX}$, $\varrho_{GY}$,
$\varrho_{XB}$ and  $\varrho_{YB}$ have a decoherence rate of
$\beta=1/T_2$, which includes the dephasing caused by the finite
lifetime. Note that $T_2$ is typically called dephasing time. The
coherence between ground and biexciton state $\varrho_{GB}$ decays
with the rate $\beta_B$ and the coherence between the single exciton
states $\varrho_{XY}$ is subject to decoherence with the rate
$\beta_{XY}$. Details on the Hamiltonian and the equations of motion
can be found in the Methods section. To compare experimental data
and theoretical predictions, we fitted the analytical formulas to
the data using independently determined FSS $\delta$, the biexciton
binding energy $\Delta$ and the decay and dephasing rates as fitting
parameters. The retrieved decay and decoherence times correspond to
the values used for the theoretical curves.

A crucial input into further results are pulse areas $\theta_i$.
When varying the latter, the QD exciton undergoes Rabi
rotations~\cite{PattonPRL05,ReiterJPC14}. In the two-level system a
pulse area of $\pi$ is attained when a complete transition from the
ground state to the exciton state takes place. In FWM, where the
coherence is probed, the signal is expected to follow a
$|\sin\left(\theta_{1}\right)|$ dependence, where the first maximum
is at $\pi/2$  when the polarization is maximal. Thus, it is a
prerequisite to determine the relation between the measured driving
intensities $P_{i}=E^2_i$, to their pulse areas
$\theta_{i}=\int\mu|E_i(t)|{\rm d}t \sim \sqrt{P_i}$. To restrict
ourselves to a two-level system we use circularly polarized light to
excite and probe the system. In Fig.~\ref{fig:basic}\,b, we plot the
FWM amplitude of the $GX$ transition against $\sqrt{P_{1}}$, while
the pulse intensity of the second pulse is fixed at $P_2=1$~\textmu
W. The maximum lies at $P_1=0.25$~\textmu W (corresponding to
approximately $10^3$ photons per pulse $E_1$), which we identify as
$\theta_1=\pi/2$ pulse.

\subsection{Classification of Quantum beats}
\label{sec:basic}

\begin{figure}[t]
\centering
\includegraphics[width=\columnwidth]{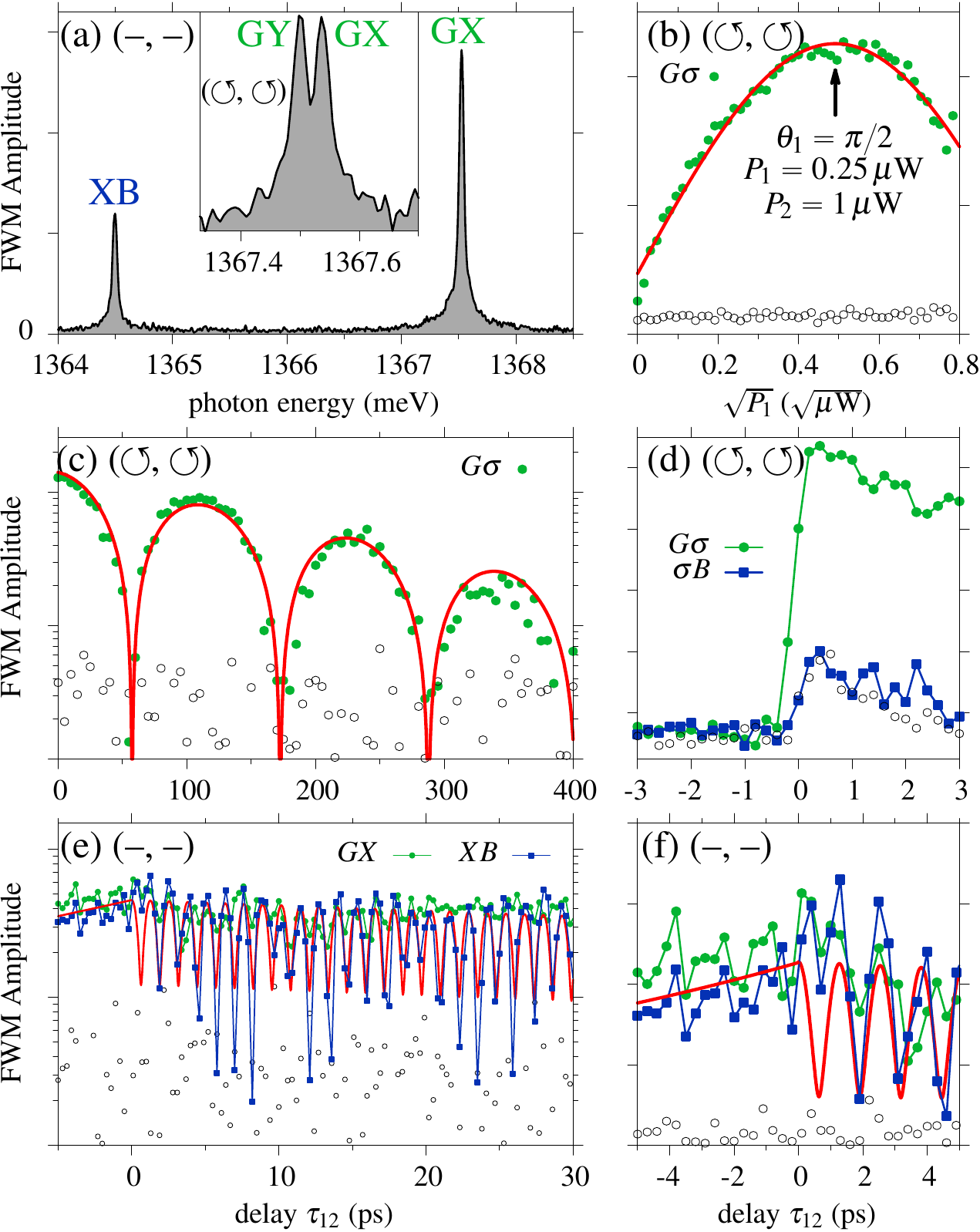}
\caption{{\bf Quantum beats.} ({\bf a}) Two-pulse FWM spectra of a
single QD for co-linear polarization. Inset: Zoom in on the exciton
line for co-circular polarization. ({\bf b}) FWM signal as function
of excitation intensity $\sqrt{P_1}$. ({\bf c}) FWM signal as
function of time delay $\tau_{12}$ for co-circularly polarized
excitation. ({\bf d}) Zoom-in of ({\bf c}) around $\tau_{12}=0$.
({\bf e}) and ({\bf f}) FWM signal as function of $\tau_{12}$ for
linear polarization with $\alpha=0^{\circ}$.  Green dots and blue
squares: experimental data; gray circles: indication of noise level;
red lines: theoretical calculation. \label{fig:basic}}
\end{figure}

The FSS $\delta$ and the BBE $\Delta$ both give rise to quantum
beats, which are visible in different FWM signals as presented in
Fig.~\ref{fig:basic}. In the spectrum the energies $\delta$ and
$\Delta$ can be determined as the difference between the respective
spectral lines. Figure~\ref{fig:basic}\,a shows a single QD spectrum
obtained from co-linearly $(\polho,\polho)$ polarized excitation.
$GX$ and $XB$ line are clearly separated by $\Delta\approx 3$~meV.
The inset shows a FWM spectrum obtained from co-circularly
$(\polcp,\polcp)$ polarized excitation, where we can see the FSS
between the two single excitons with $\delta\approx 38$~\textmu eV.

In the time domain, the period of the beatings is related to the
energy via $T_{\delta}=2\pi\hbar/\delta$ and
$T_{\Delta}=2\pi\hbar/\Delta$. The FSS-induced quantum beat can best
be seen for cocircularly polarized excitation, where no biexciton is
excited as presented in Fig.~\ref{fig:basic}\,c. The laser pulses
excite the $G\sigma$ transition, which is a linear combination of
$X$ and $Y$. When evolving in time, the $G\sigma$-coherence
oscillates with the FSS. Indeed, the FWM amplitude which is a
measure for the $G\sigma$-coherence as function of the delay
$\tau_{12}$ displays pronounced quantum beats and an exponential
decay~\cite{LangbeinPRB04, PattonPRB06, KasprzakPRB08,
KasprzakNJP13}, as reproduced by the theoretical calculations. The
corresponding equation for the FWM signal $\cal{S}_{\nu\nu^\prime}$
is given by \be \label{eq:circ}
    {\cal{S}}_{G\sigma} \propto \exp\left(-{\frac{\tau_{12}}{ T_{2}}}\right)\sqrt{\cos\left(\frac{\delta}{\hbar}\tau_{12}\right)+1}.
\ee The pulse area only determines the strength of the signal. Note,
that for a pulse area being a multiple of $\pi$, no coherence
$\varrho_{G \sigma}$ is excited and the FWM amplitude is zero. From
Fig.~\ref{fig:basic}\,c we extract the period of the FSS-induced
beating to $T_{\delta}=115$~ps, which corresponds to
$\delta=36~$\textmu eV in good agreement with the spectral
measurement in Fig.~\ref{fig:basic}\,a. We retrieve a dephasing time
of $T_2=200$~ps. Under circularly polarized excitation the FWM
signal of the $\sigma B$ transition is suppressed (cf.
Fig.~\ref{fig:schema}\,a) and accordingly there is no signal at the
$\sigma B$ transition. When we zoom into the coherence dynamics at
delays close to $\tau_{12}=0$ in Fig.~\ref{fig:basic}\,d, we confirm
that the $\sigma B$ transition does not exceed the background level,
which for $\tau_{12}>0$ is enhanced by the broadband emission of
acoustic phonons generated by $G\sigma$. For negative time delays
the FWM signal is zero since there is no two-photon coherence under
circularly polarized excitation~\cite{KasprzakNJP13}.

To determine the BBE $\Delta$ in the time domain, we excite with
co-linear polarization along the QD axis, such that the system can
be reduced to the three levels $GXB$ (cf. Fig.~\ref{fig:schema}\,b).
In this case, there is no FSS-induced quantum beat between the
excitons, because we drive the energy eigenstate $X$, while the
transitions $GY$ and $YB$ are not excited. This gives us the
opportunity to solely study the BBE-induced quantum beat. The
dynamics of the FWM signal is presented in Fig.~\ref{fig:basic}\,e,
where the FWM signals of the $GX$ and the $XB$ transition are shown
as function of the time delay $\tau_{12}$. A strong oscillation of
the $XB$ signal with a period of $T_{\Delta}=1.27$~ps is found,
corresponding to a BBE of $\Delta=3.25$~meV. These values are in
good agreement with the spectral measurements in
Fig.~\ref{fig:schema}\,a. The oscillation is also seen in the $GX$
signal, but with a much weaker amplitude. The oscillation gets more
pronounced with increasing pulse area. The dynamics can be described
analytically with our formalism~\cite{KrugelPRB07}. The general
formula for arbitrary pulse areas is rather lengthy, therefore we
give the equations which have been evaluated for our parameters,
namely the polarization angle $\alpha=0^\circ$ and pulse area. With
$\theta_1=3\pi/5$ we obtain
\begin{subequations}
\label{eq:lin}
\bea
    {\cal{S}}_{GX} \propto \exp\left(-{\frac{\tau_{12}}{ T_{2}}}\right)\sqrt{4.36 + \cos\left(\frac{\Delta}{\hbar}\tau_{12}\right)},\\
    {\cal{S}}_{XB} \propto \exp\left(-{\frac{\tau_{12}}{ T_{2}}}\right)\sqrt{1.15 + \cos\left(\frac{\Delta}{\hbar}\tau_{12}\right)}. \label{eq:BBE_XB}
\eea
\end{subequations}
Note that the sign in front of the cosine is negative for pulse
areas $\theta_1 < \pi/2$ and positive for pulse areas $\theta_1 >
\pi/2$. For clarity we only show Eq.~\eqref{eq:BBE_XB} in
Fig.~\ref{fig:basic}\,e and f.

For negative time delays a two-photon coherence $\varrho_{GB}$
between ground state and biexciton is generated by the pulse $E_2$,
which arrives first at the QD. Due to the reversed ordering this
coherence is then transferred into the FWM signal by $E_1$ showing a
decay in both $GX$ and $XB$ transition according to \be
\label{eq:neg}
    {\cal{S}}^{\tau<0}_{GX,XB} \propto e^{-\beta_B \tau_{12}}.
\ee This is in agreement with the signals shown in
Fig.~\ref{fig:basic}\,e and f, where a fast decay for increasing
negative time delays is observed.

Because the BBE-induced oscillation is much faster than the
FSS-induced beating, to resolve them a much longer measurement time
is required. Thus, in the following, we performed measurements such
that the BBE oscillation is not resolved. We further note that the
results in the following Secs.~\ref{sec:coh} and \ref{sec:pop} were
obtained from different QDs, which exhibit the same behavior but
with different time constants.

\subsection{Polarisation angle dependent coherence dynamics}
\label{sec:coh}

\begin{figure}[htbp]
\centering
\includegraphics[width=\columnwidth]{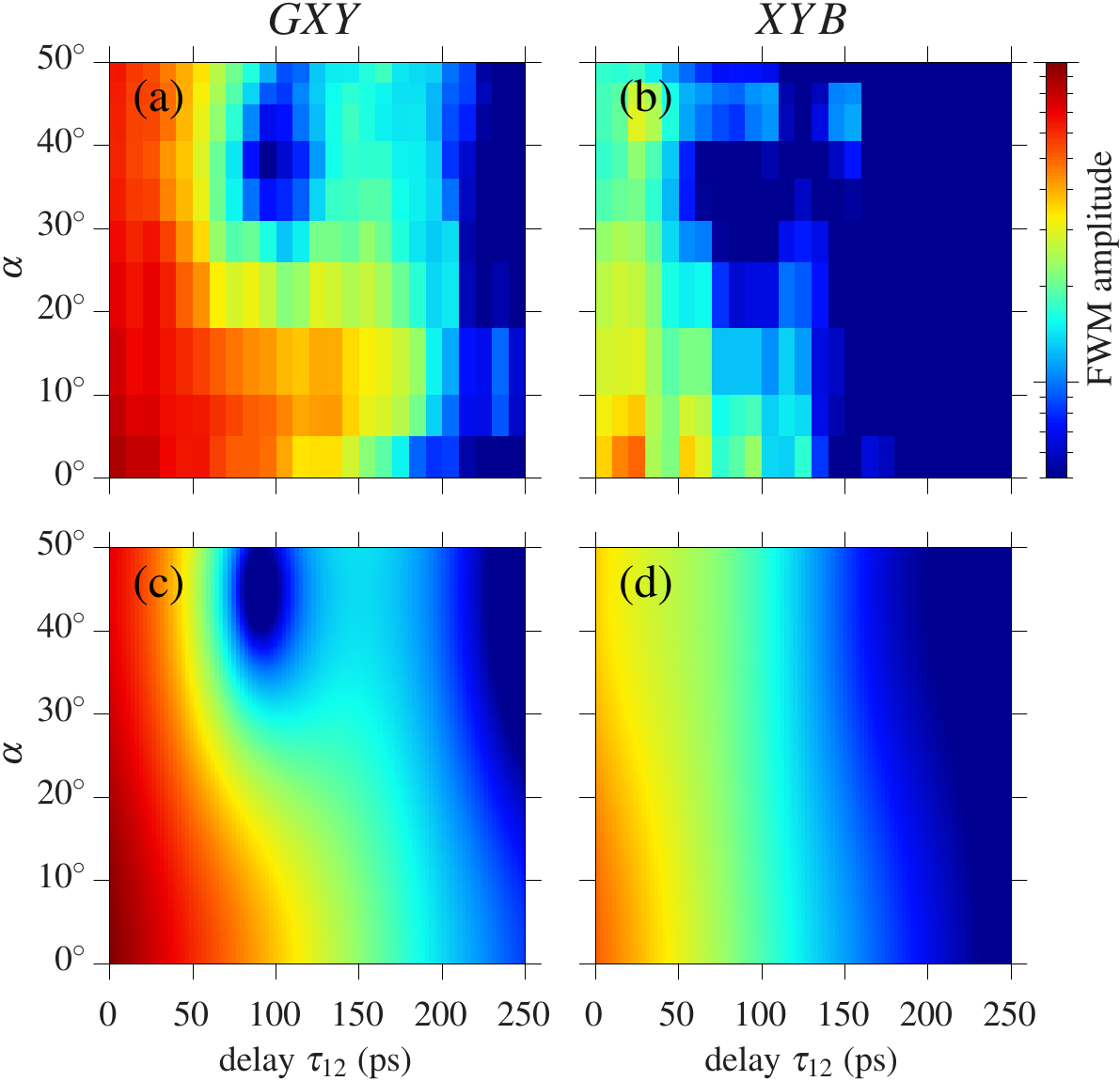}
\caption{{\bf Coherence dynamics for co-linear excitation.} ({\bf
a,b}) Measurements and ({\bf c,d}) calculations of the coherence
dynamics as function of delay $\tau_{12}$ and polarization angle
$\alpha$ for co-linear excitation for ({\bf a,c}) the $GXY$
transition and ({\bf b,d}) the $XYB$ transition.  \label{fig:3d}}
\end{figure}

As soon as the polarization of the exciting light field deviates from the strictly co-circular ($\polcp$,$\polcm$)  or co-linear polarization with $\polve$ or $\polho$, where the QD system can be reduced to a two- or three-level system, respectively, all four transitions are excited and the dynamics is governed by a mixture of the FSS-induced quantum beat and dephasing. One important control parameter is the polarization angle $\alpha$ of the exciting laser fields. In Fig.~\ref{fig:3d} the FWM signals as function of the time delay $\tau_{12}$ and the polarization angle $(\alpha,\alpha)$ for a co-linear two pulse excitation are shown. The pulse area is $\theta_1=\pi/5$. There is an excellent agreement between the measured signals in Fig.~\ref{fig:3d} (upper row) and the theoretical calculations in Fig.~\ref{fig:3d}~(lower row).

As explained above, for $\alpha=0^\circ$ no FSS-induced beat is
observed neither in the $GXY$ nor the $XYB$ signal, as the
excitation is along the QD axis. Instead, we just observe a decay
with the dephasing time $T_2$, which we here extract to
$T_2=200$~ps. In contrast, for $\alpha=45^{\circ}$ the FSS-induced
beat is maximal in the $GXY$ signal. For this angle, an equal
superposition of $X$ and $Y$ is excited invoking the FSS-induced
quantum beat with a period of $T_\delta=180$~ps. For intermediate
angles between $\alpha=0^{\circ}$ and $45^{\circ}$ there is a smooth
transition with a less pronounced beating structure. Such angle
dependent measurements can be used to determine the absolute axes
($x,\,y$) of the linearly polarized transitions $GX$ and $GY$. In
contrast to the circularly polarized excitation, for linear
excitation also the $XB$ transitions is driven by the laser field
and we see a finite $XB$ signal. However, for such small pulse areas
the $XYB$ signal is very weak, because the biexciton transition is
barely excited. Accordingly no quantum beats are seen in the $XB$
signal. If the excitation were at higher pulse areas, also for the
$XYB$ signal we would obtain FSS-induced quantum beats.

\begin{figure}[h]
\centering
\includegraphics[width=1.02\columnwidth]{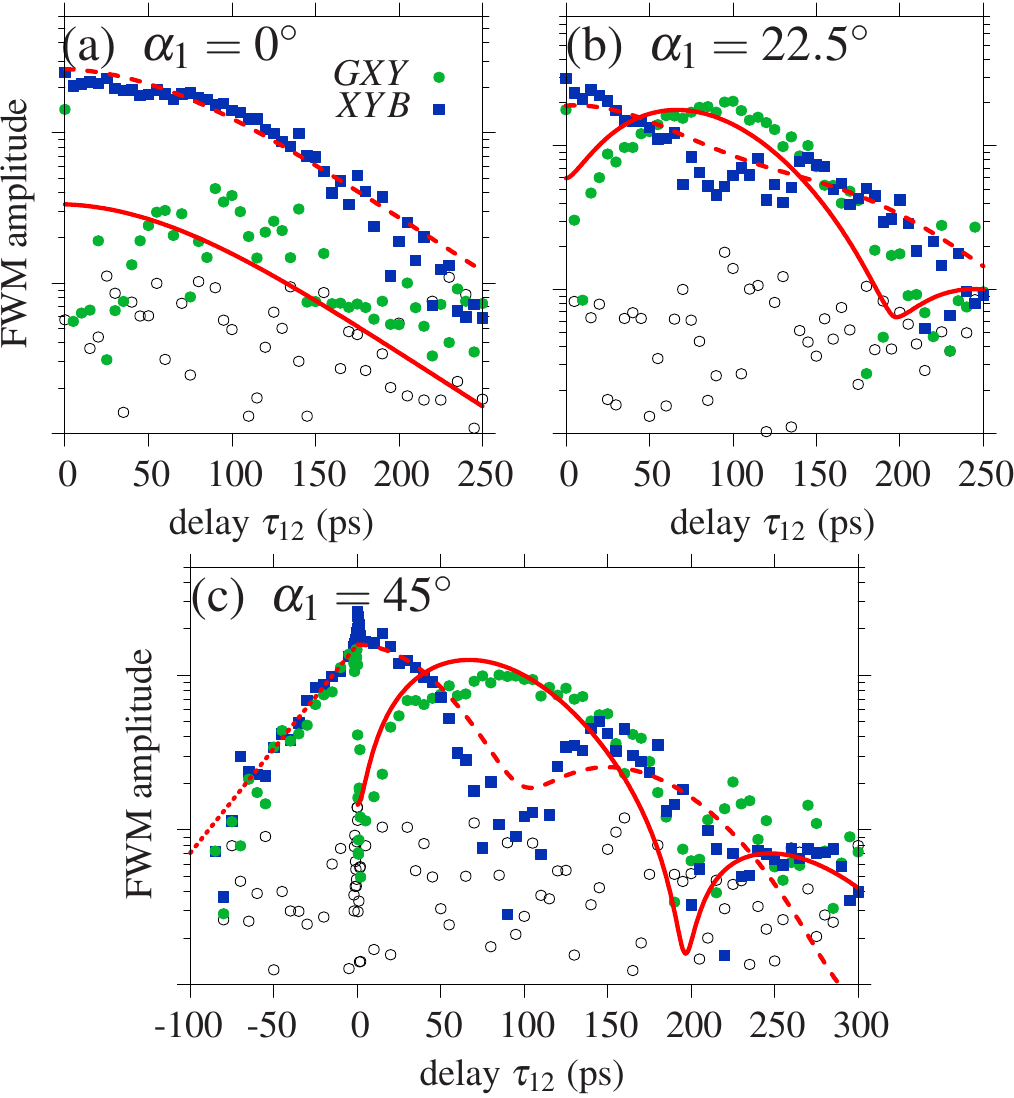}
\caption{{\bf Coherence dynamics for cross-linear excitation.}
Two-pulse FWM amplitude of the coherence dynamics of $GXY$ and $XYB$
transition for cross-polarized excitation with
$\alpha_2=\alpha_1+90^{\circ}$ and ({\bf a}) $\alpha_1=0^{\circ}$,
({\bf b}) $\alpha_1=22.5^{\circ}$ and ({\bf c})
$\alpha_1=45^{\circ}$. Green dots and blue squares:\,experimental
data of $GXY$ and $XYB$, respectively. The background level is
indicated by gray circles. Theoretical predictions are given by red
lines. \label{fig:cross}}
\end{figure}

We also analyze the angle dependence for cross-polarized driving in
Fig.~\ref{fig:cross}, where the angle of the second pulse is
perpendicular to the first pulse, i.e., $\alpha_2 = \alpha_1 +
90^{\circ}$. Here the pulse area is stronger with $\theta_1=3\pi/8$.
For $\alpha_1=0^{\circ}$ the FWM is primarily driven at $XB$ via
so-called Raman coherence induced between $X$ and $Y$ excitons in
the second-order. Conversely, the $GX$ almost vanishes, as shown in
Fig.~\ref{fig:cross}\,a. Additionally, we see a bending of the curve
for initial $\tau_{12}$. This bending is attributed to inhomogeneous
broadening caused by charge fluctuations in the QD environment,
which we also account for in our model (see Methods section). For
$\alpha=45^{\circ}$ shown in Fig.~\ref{fig:cross}\,c a pronounced
oscillation is observed in both $GXY$ and $XYB$. Having a closer
look at small delays we find that the $XYB$ signal starts with a
maximum, while the $GXY$ signal has a minimum attaining the noise
level at $\tau_{12}=0$. For the intermediate angle
$\alpha=22.5^{\circ}$ the signals are almost equally strong and both
exhibit pronounced FSS-induced beating. For $\alpha_1=45^{\circ}$,
we measured the FWM signal at negative delays $\tau_{12}$, i.e.,
$E_2$ comes before $E_1$, presented in Fig.~\ref{fig:cross}\,c.
Similar to the case of co-linear excitation with $\alpha=0^\circ$,
$E_2$ excites a two-photon coherence between ground and biexciton
state. The corresponding decay with the decoherence rate $\beta_B$
is seen and can also be described by Eq.~\eqref{eq:neg}. For the
decay rate $\beta_B$ we find $T_B=1/\beta_B=91$~ps.

\subsection{Population dynamics}
\label{sec:pop}

\begin{figure}[htbp]
\centering
\includegraphics[width=\columnwidth]{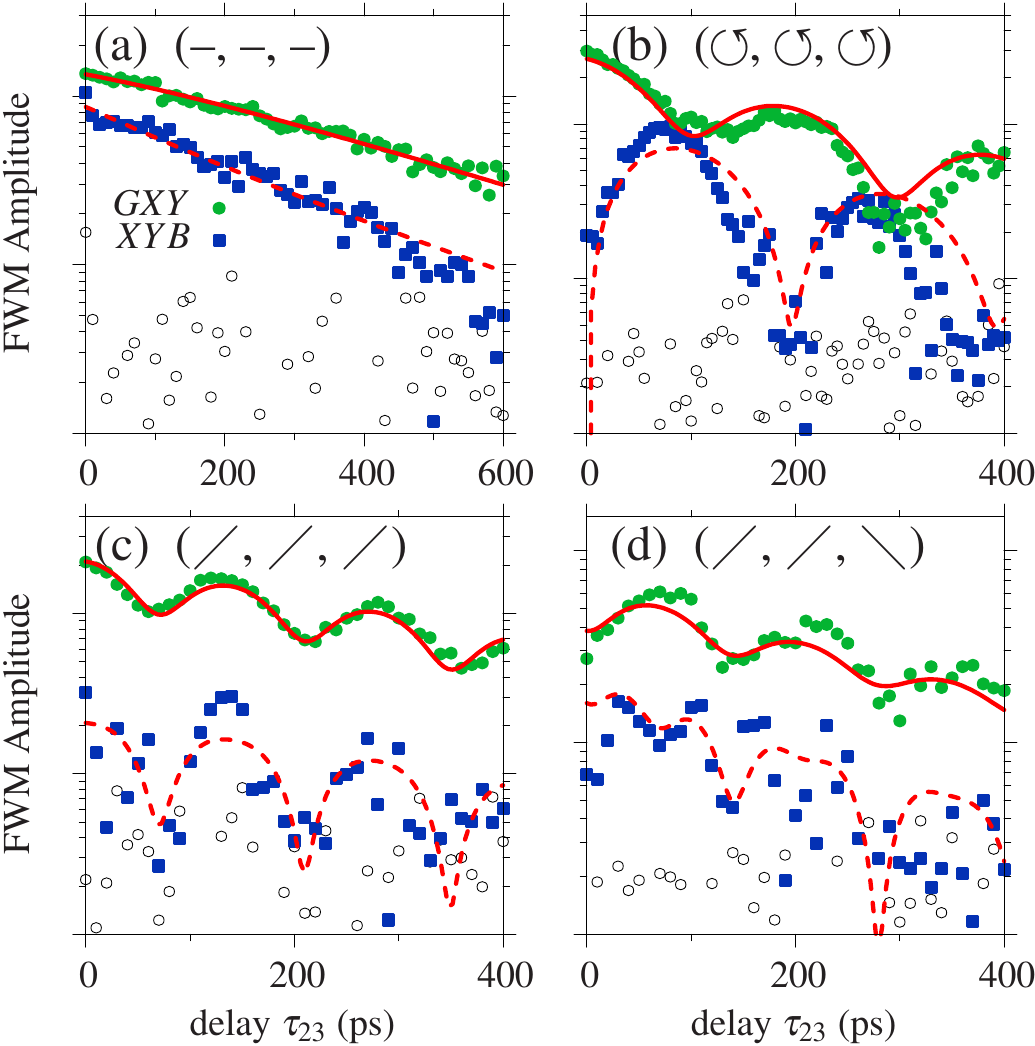}
\caption{{\bf Population dynamics.} Three-pulse FWM amplitude
displaying population dynamics of $GXY$ and $XYB$ for different
polarization states of $E_{1,2,3}$ ({\bf a})
$(\polho,\polho,\polho)$, ({\bf b}) $(\polcp,\polcp,\polcp)$, ({\bf
c}) $(\poldu,\poldu,\poldu)$ and ({\bf d}) $(\poldu,\poldu,\poldd)$.
Green dots and blue squares: experimental data of $GXY$ and $XYB$,
respectively; gray circles: indication of noise level; red lines:
theoretical calculation. \label{fig:pop}}
\end{figure}

Next, we focus on population dynamics. For a two-level system
excited by three pulses, the first pulse creates a coherence, the
second pulse creates a population and the third pulse induces the
FWM signal which is probed. Therefore, information on the decay rate
$\gamma$ can be gained in such an experiment. In
Fig.~\ref{fig:pop}\,a we show the $GXY$ and $XYB$ signal for the
three pulse excitation with $\tau_{12}=0.5$~ps as function of time
delay $\tau_{23}$. The analytical formula shows that both signals
decay without any oscillation including an exponential decay with
$\gamma$ and $2\gamma$. Instead of showing the general equation, we
again evaluate the formulas for the explicit pulse area
$\theta=0.85\pi$ yielding
\begin{subequations}
\begin{eqnarray}
\mathcal S_{GXY} &\propto & e^{-\gamma\tau_{23}} - 0.29 e^{-2\gamma\tau_{23}},\\
\mathcal S_{XYB} &\propto & e^{-\gamma\tau_{23}} + 0.76
e^{-2\gamma\tau_{23}}.
\end{eqnarray}
\end{subequations}
From the exponential decay of the FWM amplitude we retrieve a
lifetime of $T_{1}=1/\gamma=333$~ps.

When we choose a different excitation polarization, we find an
oscillatory FWM signal in Fig.~\ref{fig:pop}\,b-d. Let us start with
circularly polarized excitation $(\polcp,\polcp,\polcp)$ shown in
Fig.~\ref{fig:pop}\,b. The first two pulses create a population of
the $\sigma^{+}$-exciton. From this state the third pulse cannot
excite the biexciton. Consequently, at $\tau_{23}=0$ the signal
$\mathcal{S}_{XYB}$ is zero. Remember that $\sigma^{+}$ is a linear
combination of the $X$ and $Y$ exciton. After the first two pulses,
the populations and the polarization $\rho_{XY}$ between $X$ and $Y$
contribute to the FWM. The latter oscillates with the FSS $\delta$,
which is directly reflected in the $GXY$ signal. The period is
$T_{\delta}=196$~ps corresponding to a FSS $\delta=21$~\textmu eV.
The quite different value in FSS compared to the QD used in
Sec.~\ref{sec:basic} shows that the FSS depends sensitively on the
QD under examination. When the oscillation sets in, the exciton
changes its character from $\sigma^{+}$ to $\sigma^{-}$ due to the
oscillating polarization $\varrho_{XY}$. When the system is in the
$\sigma^{-}$ exciton, the biexciton can be excited and the $XYB$
signal has a maximum coinciding with $GXY$ having a minimum. The
equations for these dynamics are given by
\begin{subequations}
\bea
&&\mathcal S_{G\sigma}  = \mathcal S_{GXY} \\ \nonumber
&&\propto  \sqrt{6\cos\left(\frac{\delta \tau_{23}}{\hbar}\right)e^{-(\gamma+\beta_{XY})\tau_{23}} + 9e^{-2\gamma\tau_{23}} + e^{-2\beta_{XY}\tau_{23}}},\\
&&\mathcal S_{\sigma B} = \mathcal S_{XYB} \\ \nonumber &&\propto
\sqrt{2\cos\left(\frac{\delta\tau_{23}}{\hbar}\right)e^{-(\gamma+\beta_{XY})\tau_{23}}
- e^{-2\gamma\tau_{23}} - e^{-2\beta_{XY}\tau_{23}}}. \eea
\end{subequations}
Similar to the coherence dynamics, the pulse area only determines
the prefactor, but does not influence the time-dependence. We see an
oscillatory term with the FSS $\delta$, which is damped by the decay
rate $\beta_{XY}$ corresponding to the coherence $\varrho_{XY}$
between the single excitons.

For linearly polarized excitation with a diagonal polarization
$(\poldu,\poldu,\poldu)$ presented in Fig.~\ref{fig:pop}\,c, again a
superposition of $X$ and $Y$ with an oscillating polarization
$\varrho_{XY}$ is excited. Accordingly $GXY$ oscillates likewise
with the FSS. Note, that here a different QD was examined, for which
we find the values $T_{\delta}=140$~ps and $\delta=30$~\textmu eV.
In contrast to the case of circular excitation, for linear
excitation the biexciton can be directly addressed. Accordingly, the
$XYB$ signal starts with a maximum and then oscillates with the same
period as the $GXY$ signal. Finally, we examine the cross-polarized
excitation with $(\poldu,\poldu,\poldd)$ in Fig.~\ref{fig:pop}\,d. A
general trend is that $GXY$ for cross-linear excitation oscillates
opposite to the case of co-linear excitation. Because all
polarizations contribute to the signal, also a mixture of the $GX$
and $GY$ occurs, leading to more complex behavior in the $XYB$
signal involving higher harmonics of the beat frequency.

\section{Transitions in a Charge fluctuating QD}

\begin{figure}[htbp]
\centering
\includegraphics[width=\columnwidth]{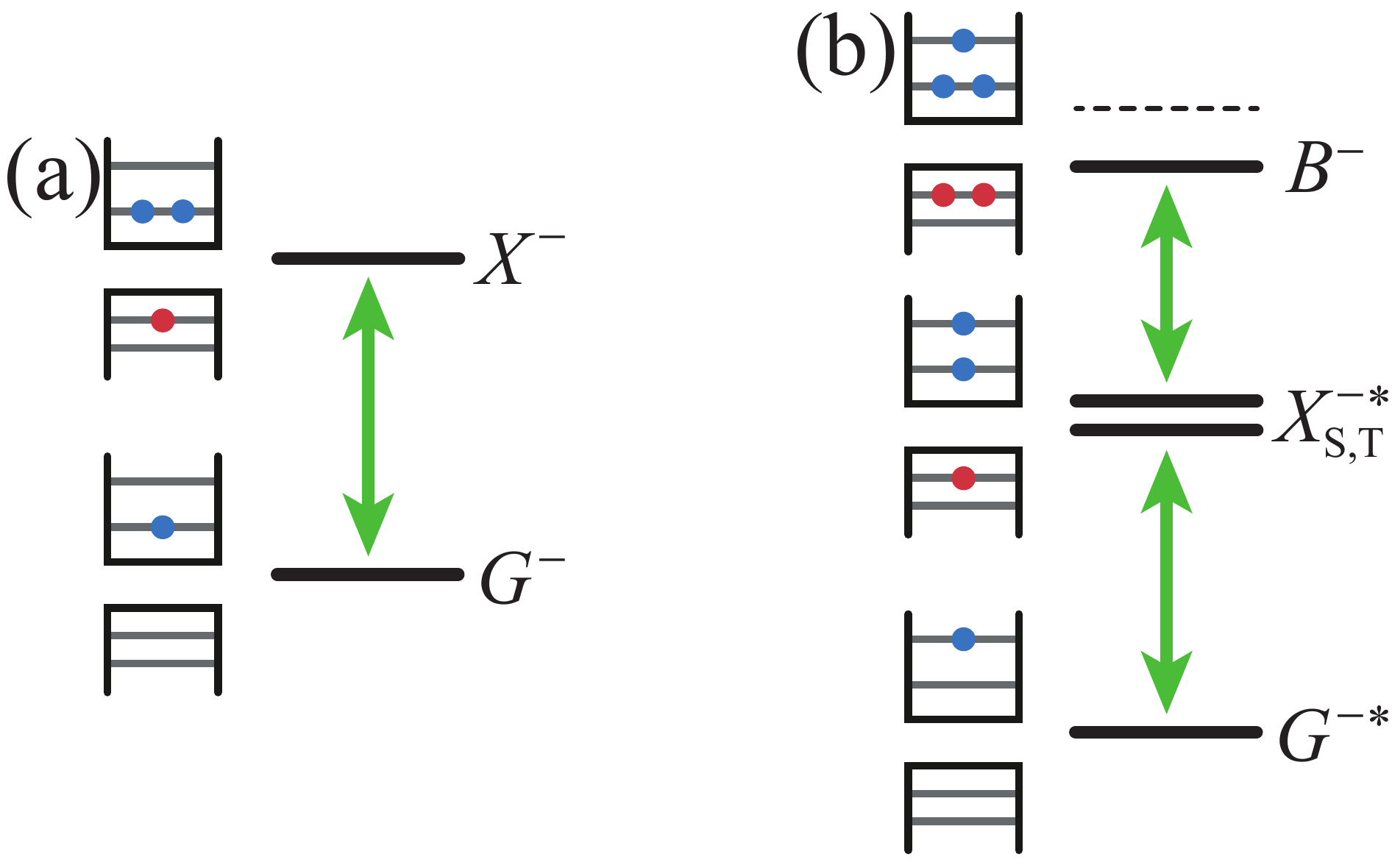}
\caption{{\bf Charged exciton complexes.} Pictographic diagram of
the ({\bf a}) negative trion system and ({\bf b}) the charged
biexciton system with the excited trion states.
\label{fig:complexes}}
\end{figure}

\begin{figure}[htbp]
\centering
\includegraphics[width=\columnwidth]{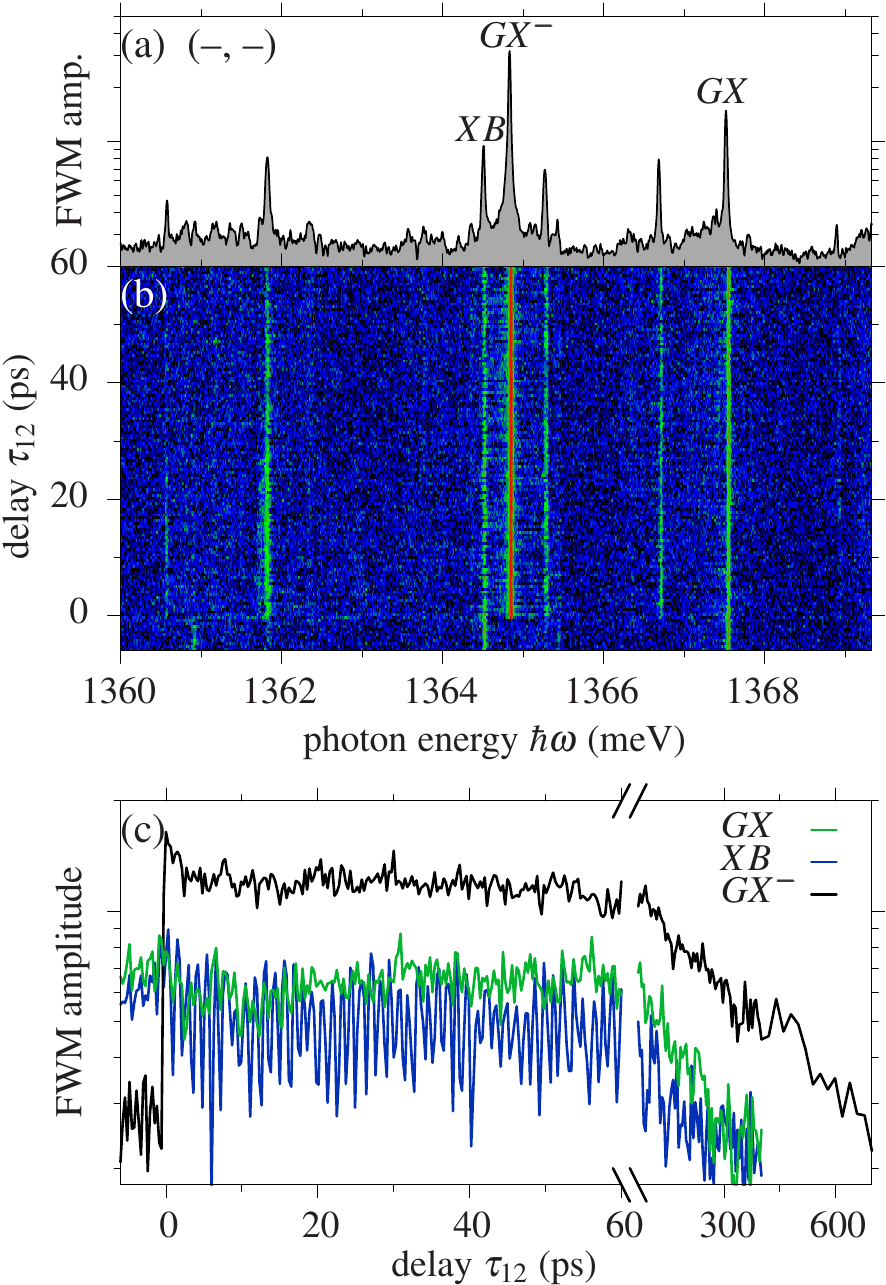}
\caption{{\bf FWM of a charge fluctuating QD.} ({\bf a}) Two-pulse
FWM spectrum of a QD showing with ({\bf b}) the corresponding
dynamical behavior in $\tau_{12}$. ({\bf c}) FWM signal of the
coherence dynamics for the $GX$, $XB$ and $GX^-$ transition.
\label{fig:charged}}
\end{figure}

The charge state of a QD can fluctuate~\cite{KuhlmanNatPhys13} over
timescales several orders of magnitude faster than the integration
time, which is in the 1-100 second range. Thus, not only the neutral
exciton and biexciton appear in the FWM spectrum, but an assortment
of different neutral and charged exciton transitions. One example of
such a spectrum integrated over the delay time $\tau_{12}$ is shown
in Fig.~\ref{fig:charged}\,a. The spectrum shows a variety of lines
spread over a few meV. Because our sample is $n$-doped, we
predominantly find dots which are initially charged with a single
electron having the ground state $G^{-}$. When such a dot is excited
the negatively charged exciton (trion) $X^{-}$ is generated via the
transition $GX^{-}$ as depicted in Fig.~\ref{fig:complexes}\,a. Due
to the doping, $GX^{-}$ has a high intensity and can be identified
as the line at $1364.8$~meV in Fig.~\ref{fig:charged}a.

To identify the $GX$ and $XB$ transition of the neutral exciton, we
look at the delay dependence of the coherence. The temporal behavior
of all states is shown in Fig.~\ref{fig:charged}\,b, while the
dynamics of the most fundamental lines, namely the $GX$, $XB$ and
$GX^-$ line is displayed in more detail in
Fig.~\ref{fig:charged}\,c. From the previous results, we know that
the neutral $GX$ and $XB$ create a FWM signal for $\tau_{12}<0$ via
the two-photon coherence. In addition we checked that these
transitions obey the corresponding polarization selection rules of
GXYB (not shown), as detailed in the previous section~\ref{sec:coh}.
Thus we identify $GX$ at $1367.5$~meV and $XB$ at $1364.5$~meV. In
contrast, negative trion lack FWM for $\tau_{12}<0$, since the
corresponding charged biexciton is spectrally too far to be
excited\,\cite{KordianoSST14}. This is confirmed by looking at
$GX^-$ in Fig.~\ref{fig:charged}\,b and c, where $GX^-$ is zero for
$\tau_{12}<0$. For positive time delays $\tau_{12}>0$ we can
discriminate between the transitions by the quantum beats. We see a
strong quantum beat in the $XB$-transition induced by the BBE as in
Fig.~\ref{fig:basic}\,e with a period $T_{\Delta}=1.25$~ps. The
BBE-induced oscillation survives on a long time scale of several
hundreds of ps. The charged exciton transition $GX^-$ does not show
any dynamical behavior apart from a decay. This is expected, since
$G^{-}$ and $X^-$ form a two-level system, depicted in
Fig.~\ref{fig:complexes}\,a, where only the dephasing alters the
signal. We find a dephasing time of $T_2=270$~ps.

\begin{figure}[htbp]
\centering
\includegraphics[width=\columnwidth]{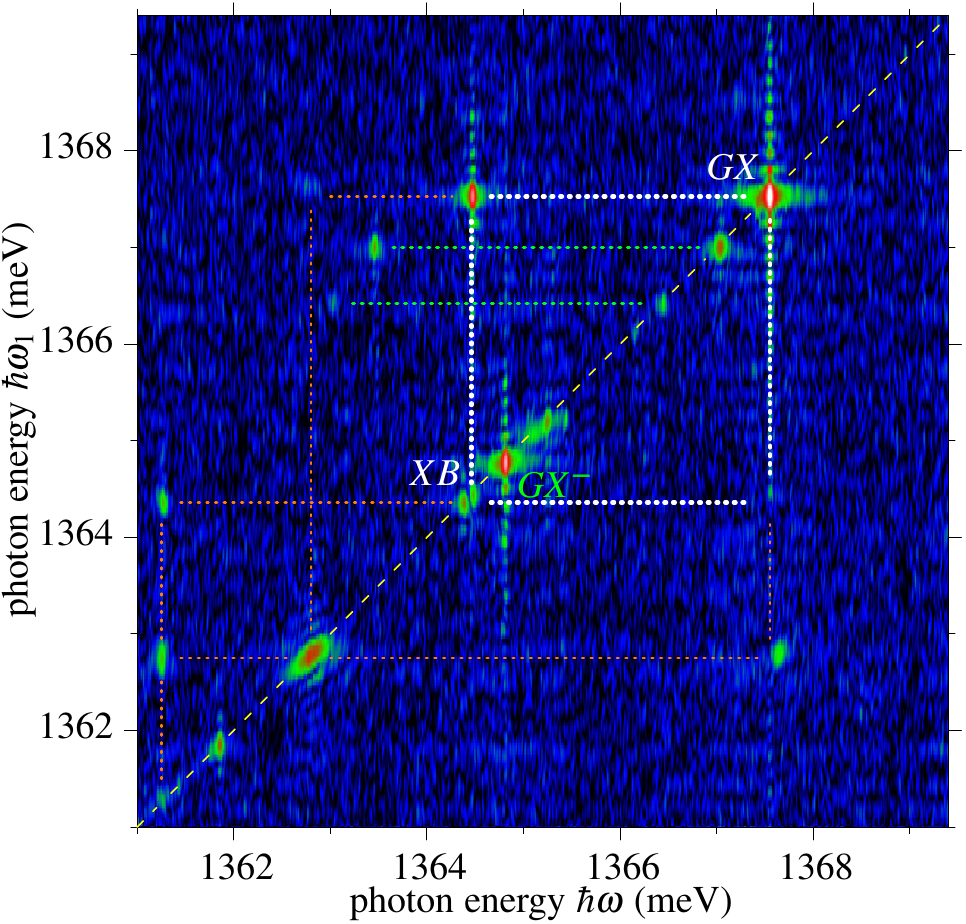}
\caption{{\bf Two-dimensional four-wave-mixing.} 2D FWM spectroscopy
map,
 revealing coherent couplings between exciton complexes
generated by a charge fluctuating QD. \label{fig:2D}}
\end{figure}

More information about the different exciton complexes can be gained
from 2D spectral FWM maps. These maps are obtained by a 2D Fourier
transform with respect to the real time $t$ (horizontal axis) and
the delay time $\tau_{12}$ between the pulses (vertical axis). While
the transform with respect to $t$ is assured by the spectrometer,
the one with respect to $\tau_{12}$ requires adjusting the phase
evolution for different delays~\cite{KasprzakNPho11} by implementing
``the guiding star approach". As such a reference transition we
choose the uncoupled $GX^-$. 2D FWM correlates resonances active in
the first-order absorption, $\omega_1$ (vertical axis), with the FWM
ones, $\omega$ (horizontal axis). The spectrum can be regained from
the 2D map by integrating over the y-axis. An advantage of the 2D
spectrum is the possibility to detect couplings between different
exciton transitions, which appear as off-diagonal peaks. The neutral
exciton cast should be correlated among itself and strictly
separated from charged configurations. An example of a 2D map from
the same dot as in Fig.~\ref{fig:charged}, but for a higher
excitation power, is shown in Fig.~\ref{fig:2D}. Let us focus on the
neutral exciton complexes first: There are two points on the
diagonal line at  $1367.5$~meV and at $1364.5$~meV, which can be
clearly identified as $GX$ and $XB$ transition, respectively. From
the dynamics in Fig.~\ref{fig:charged}\,c, we have seen that $XB$
shows a BBE-induced beating. In the 2D map, this is seen by a
off-diagonal peak at ($1364.5$~meV, $1367.5$~meV). We can identify a
couple of other peaks, which are connected with the neutral exciton
complex. These are marked by orange lines. These could stem from
exciton or biexciton complexes, where an electron or a hole is in a
p-shell, or from higher order non-linear processes~\cite{DaiPRL12}.

For the charged excitons, we also see corresponding diagonal points.
The strongest signal at  $1364.8$~meV belongs to the $GX^{-}$
transition and is not connected to other transitions via
off-diagonal peaks. This confirms, that $GX-$ does not couple to
other transitions and $G^{-}$ and $X^{-}$ can be modeled as a
two-level system. The coupling to the negatively charged biexciton
$B^{-}$ is unlikely. Remember that $B^{-}$ consists of three
electrons, two in the s-shell and one in the p-shell, and two
s-shell holes as depicted in Fig.~\ref{fig:complexes}\,b. In
principle, it is possible to have the transition from the trion
$X^{-}$ by exciting an exciton in the p-shell into the excited
biexciton. However, the p-shell exciton is energetically far away,
such that this transition is not covered by the laser pulse.

On the other hand, the charged biexciton $B^{-}$ can decay by
recombination of an s-shell exciton into the excited trion state
$X^{-*}_{\rm S,T}$. In the excited trion state, one electron is in
the s-shell and one in the p-shell. Due to the exchange interaction
between the electrons the excited trion splits up into a singlet
$X^{-\ast}_{\rm S}$ and a triplet $X^{-\ast}_{\rm T}$ state which
are typically separated by a few
meV~\cite{UrbaszekPRL03,AkimovPRB05}. From the excited trion, again
a recombination of an s-shell exciton can take place resulting in
the excited charged ground state $G^{-\ast}$. Such a three-level
system can give rise to off-diagonal peaks reflecting the coupling
between the states. In the 2D map in Fig.~\ref{fig:2D} we see two
such diagonal peaks at $1367$~meV and $1366.5$~meV. These have
corresponding off-diagonals connected by green lines. It is highly
likely that these are charged exciton complexes, because they are
not connected to any neutral transition, and they can probably be
identified with singlet-triplet transitions. However, because we
cannot exclude that these peaks correspond to positively charged
states, which would agree with recent photoluminescence excitation
measurements~\cite{KordianoSST14}, we refrain from a definite
attribution. Complementary insights into the biexcitonic structure
of charged states could be gained by inferring FWM beatings at
negative delays.

\begin{figure}[htbp]
\centering
\includegraphics[width=\columnwidth]{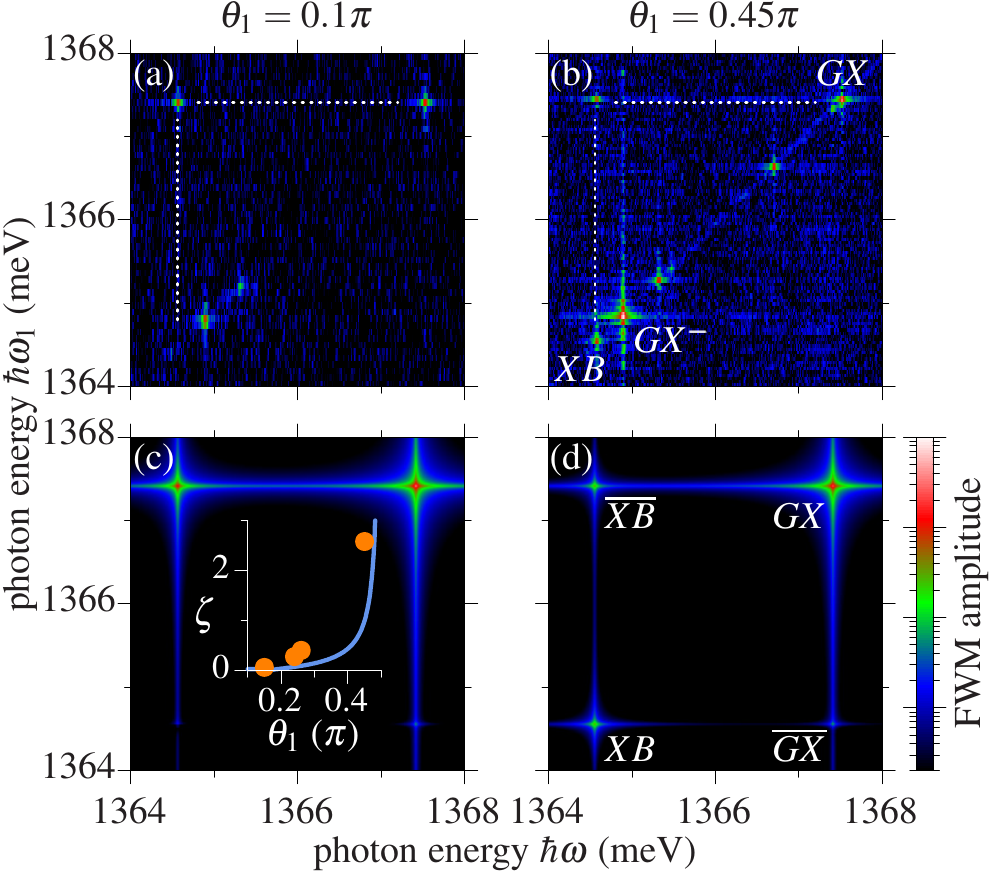}
\caption{{\bf Pulse area dependence of 2D maps.} 2D FWM spectroscopy
map of the exciton-biexciton system for an excitation with a pulse
area of ({\bf a,\,c})  $\theta_1=0.1\pi$ and ({\bf b,\,d})
$\theta_1=0.45\pi$. Upper row: experimental data, lower row:
theoretical calculations. The inset in ({\bf c}) shows the strength
of the ratio $\zeta$ of biexciton $XB$ and correlation peak
$\overline{XB}$ as function of pulse area $\theta_1$; orange dots:
experiment and blue curve: theory. \label{fig:pulsearea}}
\end{figure}

We complete the analysis of states by analyzing the dependence of
the FWM maps on the pulse area. Higher-order FWM contributions, and
hence the beating, are suppressed at lower excitation power.
Accordingly the intensities of the peaks depend crucially on the
excitation intensity. This is exemplified in
Fig.~\ref{fig:pulsearea}, where the 2D map of the $GXB$ system is
shown for two different pulse areas of $\theta_1=0.1\pi$ and
$\theta_1=0.45\pi$. The upper row shows the experimental data, while
in the lower row theoretical predictions are plotted. For low pulse
area $\theta_1=0.1\pi$, we clearly see the peak on the diagonal
corresponding to $GX$. Also the off-diagonal peak is clearly
visible, while the diagonal peak corresponding to $XB$ vanishes.
This is explained as follows: The coherence $\varrho_{XB}$ is not
excited for small pulse areas, however the second pulse $E_2$ (pulse
area twice as large) probes both transitions and, thus, the
interference peak is visible. For the pulse area $\theta_1=0.45\pi$,
which is close to $\pi/2$, the off-diagonal peak is much weaker,
while the diagonal peak for $XB$ becomes visible. The inset in
Fig.~\ref{fig:pulsearea} shows the ratio of intensities
$\zeta=I_{XB}/I_{\overline{XB}}$ between the diagonal $XB$ peak and
the off-diagonal $\overline{XB}$ peak (blue curve: theory; orange
dots: experiment), which is increasing with increasing pulse area.

\section{Conclusion}
In conclusion, we have presented a combined experimental and
theoretical study on FWM signals retrieved from single,
strongly-confined InAs QDs embedded in a low-Q semiconductor
microcavity. The experimental results are in excellent agreement
with simulations performed in a four-level system including FSS and
BBE. The latter give rise to rich and pronounced quantum beats in
the corresponding FWM signals, allowing to determine quantitative
values. From all measurements population decay and dephasing rates
were extracted. Additionally, we discussed the angle dependence of
the FWM upon co- and cross-polarized excitation. We revealed and
exploited coherences in a four-level system that are usually hidden,
specifically as regards biexciton dephasing, studied via two-photon
coherence, as well as the interplay between $X$ and $Y$ polarized
excitons induced via Raman coherence. Using 2D FWM spectroscopy we
confirmed the coupling between exciton and biexciton states and
furthermore we identified charged exciton complexes. The FWM
technique is a powerful tool to analyze coherent dynamics in
few-level systems. Employing photonic structures enhancing optical
coupling, it can directly be extended to other single photon
emitters, like NV centers in diamond~\cite{AharonovichRPP11} or
recently discovered single emitters in atomically thin
semiconductors~\cite{TonndorfOPTICA15,KoperskiNN15}, enabling to
explore coherence, reveal couplings and implement quantum control
protocols.

\section*{Methods}
\subsection*{Sample preparation and characterization}
\label{sec:sample} The MBE grown sample contains a layer of annealed
and capped InAs QDs with a nominal density of
$2.2\times10^9$~cm$^{-2}$. They are embedded in an asymmetric
GaAs/AlGaAs micro-cavity exhibiting a low quality
factor~\cite{HeNatNano13, HePRL13, MaierOptEx14, Fras15} $Q=170$,
resulting in a mode centered at 910-915~nm with a FWHM of around
10~nm. The femto-second laser pulse trains are spectrally matched
with such a large spectral window and efficiently penetrate into the
structure. Furthermore, the intra-cavity field is enhanced by a
factor of $\sqrt{Q}=13$ improving the coupling between $E_{1,2,3}$
and the electric dipole moment $\mu$ of the transition. Thus, the
resonant field required to drive the FWM is reduced by a factor
$Q^{3/2}\simeq2200$ and the signal-to-noise ratio of the
interferometrically detected FWM is amplified accordingly.

\begin{figure}[t]
\centering
\includegraphics[width=\columnwidth]{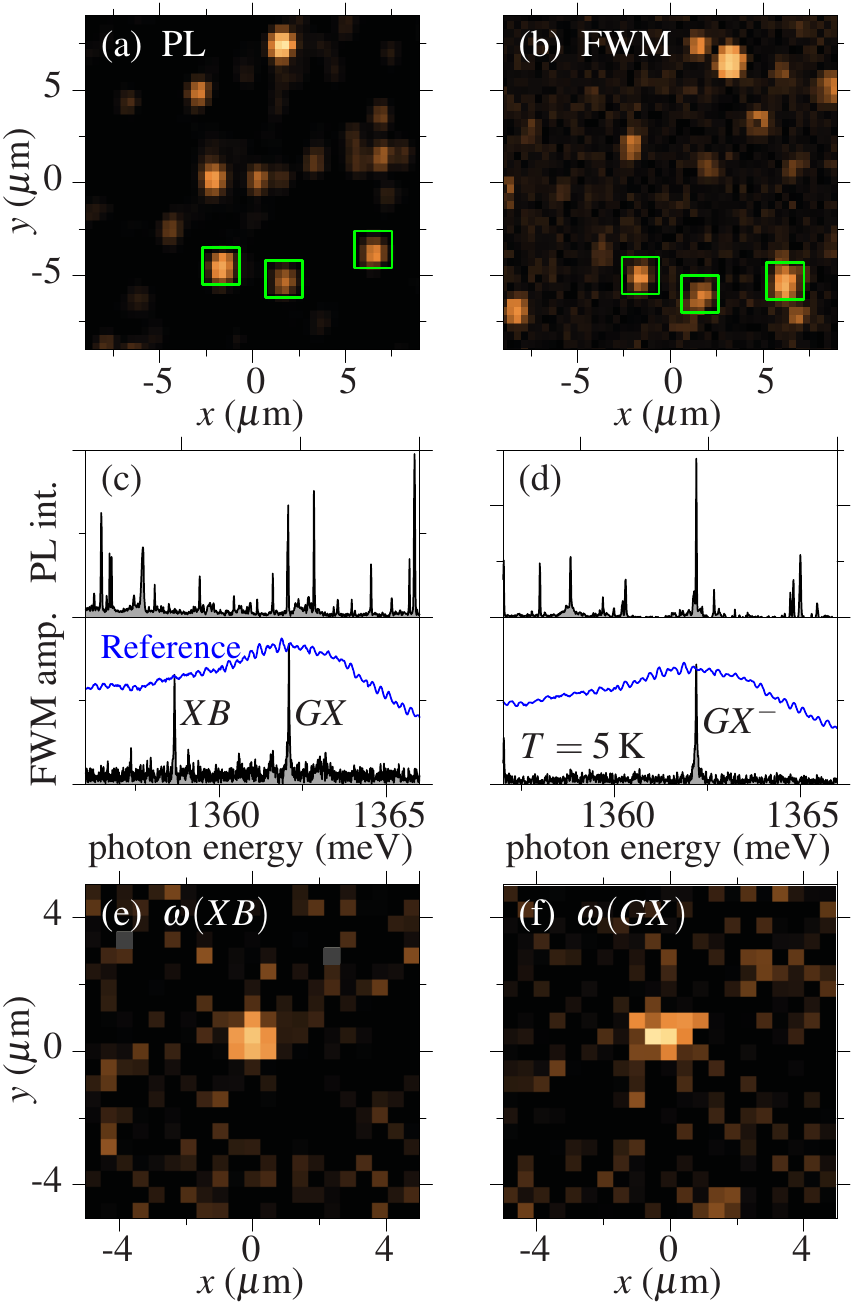}
\caption{{\bf Sample characterization.} Hyperspectral image of the
sample for ({\bf a}) PL and ({\bf b}) FWM measurements. ({\bf c,d})
PL and FWM spectra of a single QD. Blue indicates the spectrum of
the laser pulse as reference. ({\bf e,f}) Hyperspectral FWM image of
of $XB$ and $GX$ transition. \label{fig:sample}}
\end{figure}

The sample is intentionally doped with Si ($\delta$-doping with a
nominal density of $1.8\times10^{10}$~cm$^{-2}$; layer located 10~nm
below the QD plane). To identify the spatial and spectral location
of the QD transitions we perform hyperspectral
imaging~\cite{KasprzakPSSb09,KasprzakNPho11}. In
Fig.~\ref{fig:sample}\,a we present an example of such imaging
performed in a confocal micro-photoluminescence (PL) experiment.
Each bright spot corresponds to a QD emission, primarily attributed
to recombination of negative trions (GX$^-$) due to the $n$-doping.
We detect high PL counting rates on the order of $10^{5}$/sec at the
QD saturation. Such an unusually bright PL emission is attributed to
the presence of oval photonic defects on the sample
surface~\cite{MaierOptEx14, Fras15}, acting as natural
micro-lenses~\cite{GschreyNatComm15}. Additionally, the
inhomogeneous broadening due to spectral wandering is largely
reduced~\cite{HeNatNano13, Fras15} indicating an excellent
structural quality of these QDs.

The FWM hyperspectral imaging at the same sample area and spectral
range is shown in Fig.~\ref{fig:sample}\,b. The three QDs  at
$(x,y)\approx(-2~\text{\textmu m},-5~\text{\textmu m})$,
$(2~\text{\textmu m},-5~\text{\textmu m})$ and $(7~\text{\textmu
m},-4~\text{\textmu m})$ (marked with green boxes in
Fig.~\ref{fig:sample}) exhibit both PL and FWM signals and were used
to align the figures. However, other QDs show different distribution
of the peak heights in FWM as compared to the PL. This is expected
from the different properties determining the signal strength in
both measurements: in FWM the dipole moment is probed, while in PL
generally the more complex phonon-assisted carrier relaxation
combined with a capture of the exciton also lead to a signal. To
demonstrate the high spectral and spatial selectivity of the FWM
compared to the PL, Figs.~\ref{fig:sample}\,c and d compare both PL
and FWM obtained from the same sample spot, defined by the
diffraction limited size ($0.7$~\textmu m) of the excitation laser.
In Fig.~\ref{fig:sample}\,c we show a neutral exciton complex, which
is only present in few \% of the QDs. The exciton-biexciton system
is straightforwardly recognized in FWM, but it is difficult to be
identified in PL, because of a lacking $XB$ emission line.
Figure~\ref{fig:sample}\,d shows the PL and FWM spectra of the
fundamental trion line corresponding to a negatively charged QD. A
zoom-in of the spatial shape of the $XB$ and $GX$ transition is
shown in Fig.~\ref{fig:sample}\,e and f, respectively.

\subsection*{Theoretical Model}
\label{sec:theo} The Hamiltonian for circularly polarized excitation
(cf. Fig.~\ref{fig:schema}\,a) reads \bea
 H =&& \sum_{\nu} \hbar\omega_{\nu} \proj{\nu}{\nu} - \sum_{\nu,\nu^\prime} \hbar M_{\nu \nu^\prime}\proj{\nu}{\nu^\prime}+H_{\rm exc}
\eea with the basis states \be \ket{\nu} \in
\{\ket{G},\ket{\smi},\ket{\spl},\ket{B}\}\ . \ee Correspondingly the
energies  are $\hbar\omega_{G}=0$,
$\hbar\omega_{\smi}=\hbar\omega_{\spl}=\hbar\omega_{\sigma}$ and
$\hbar\omega_{B}= 2\hbar\omega_{\sigma} - \Delta$, where $\Delta$ is
the BBE. Due to the Coulomb exchange interaction the two single
exciton levels interact via \be H_{\rm exc} = \frac{\delta}{2}
\left(\proj{\smi}{\spl} +  \proj{\spl}{\smi}\right) \ee The light
field coupling \be \underline{M} = \begin{pmatrix}
0 & \Omega_{\spl}^\ast & \Omega_{\smi}^\ast & 0 \\
\Omega_{\spl} & 0 & 0 & \Omega_{\smi}^\ast \\
\Omega_{\smi} & 0 & 0 & \Omega_{\spl}^\ast \\
0 & \Omega_{\smi} & \Omega_{\spl} & 0  \\
\end{pmatrix}
\ee
describes the allowed transitions via the Rabi frequencies
\be
\Omega_{\spm} = \sum_j \frac{M_0}{\hbar} {\bf E}_j \cdot {\bf e}_{\spm}^\ast.
\ee
$M_0$ is the bulk dipole matrix element and ${\bf e}_{\spm}$ the polarization vector. The polarization of the system is given by
\bea \nonumber
 {\bf p} =&& M_0\left(\proj{0}{\spl}  + \proj{\smi}{B} \right) {\bf e}_{\spl}  \\\nonumber
  &+&M_0  \left(\proj{0}{\smi}  + \proj{\spl}{B} \right) {\bf e}_{\smi}.
\eea

This Hamiltonian can be transformed into the basis for linearly
polarized excitons (cf. Fig.~\ref{fig:schema}\,b), which are the
eigenstates of $\sum_{\nu} \hbar\omega_{\nu} \proj{\nu}{\nu}  +
H_{\rm exc}$. The transformation is calculated by \bea
&\ket{X}& = \frac{1}{\sqrt{2}}\left( \ket{\spl} + \ket{\smi} \right),\\
&\ket{Y}& = \frac{i}{\sqrt{2}}\left( \ket{\spl} - \ket{\smi} \right).
\eea
By the diagonalization the degeneracy of the single excitons is lifted and the exciton energies are $\hbar\omega_{X}=  \hbar\omega_{\sigma} - \delta/2$ and $\hbar\omega_{Y}=  \hbar\omega_{\sigma} + \delta/2$. The light field coupling changes to
\be
\underline{M} = \begin{pmatrix}
0 & \Omega_{X}^\ast & \Omega_{Y}^\ast & 0 \\
\Omega_X & 0 & 0 & \Omega_{X}^\ast \\
\Omega_Y & 0 & 0 & \Omega_{Y}^\ast \\
0 & \Omega_X & \Omega_Y & 0  \\
\end{pmatrix}
\ee
with
\be \nonumber
\Omega_X = \frac{1}{\sqrt{2}} \left( \Omega_{\spl} + \Omega_{\smi} \right),\quad \Omega_Y = \frac{i}{\sqrt{2}} \left( \Omega_{\spl} - \Omega_{\smi} \right).
\ee

The time evolution of the density matrix $\varrho$ is calculated
assuming a sum of $\delta$-pulses yielding the Rabi frequencies for
circular polarization \be \Omega_{\spm} = \sum_j
\frac{\theta_j^{\spm}}{2}\,e^{i\varphi_j^{\spm}}\delta(t-t_j) \ee
with arrival times $t_j$, pulse areas $\theta_j^{\spm}$ and phases
$\varphi_j^{\spm}$. For a pulse sequence with linear polarizations
$\alpha_j$ with respect to $X$ and pulse areas $\theta_j$ the Rabi
frequencies read \bea \nonumber \Omega_X = \ \sum_j\
\sqrt{2}\,\theta_j e^{i\phi_j} \cos(\alpha_j) \ ,\\ \nonumber
\Omega_Y = \sum_j -\sqrt{2}\,\theta_j e^{i\phi_j} \sin(\alpha_j) \ .
\eea In the case of $\delta$-pulses the time evolution of the system
can be calculated by matrix multiplication~\cite{VagovPRB02}. In
between the pulses the dynamics is given by \be \rho_{\nu
\nu^\prime}(t) = \rho_{\nu \nu^\prime}(0) e^{i\Lambda_{\nu
\nu^\prime}(t)} \ee with \bea \nonumber &\Lambda_{\nu \nu^\prime}
= & \omega_\nu - \omega_{\nu^\prime} + i\beta_{\nu \nu^\prime}\ ,\\
\nonumber &\underline{\beta} = & \begin{pmatrix}
0 & \beta & \beta & \beta_B \\
\beta & 0 & \beta_{XY} & \beta \\
\beta & \beta_{XY} & 0 & \beta \\
\beta_B & \beta & \beta & 0\\
\end{pmatrix}.
\eea
$\beta$, $\beta_{B}$ and $\beta_{XY}$ are the dephasing rates described in Sec.~\ref{sec:theo}. The decay of the exciton and biexciton is modeled by a single decay rate $\gamma$, which leads to the following equations of motion for the diagonal elements of the density matrix:
\begin{eqnarray*}
\varrho_{BB}(t) &=& \varrho_{BB}(0)e^{-2\gamma t}\\
\varrho_{XX}(t) &=& [\varrho_{XX}(0)+\varrho_{BB}(0)(1-e^{-\gamma t})]e^{-\gamma t}\\
\varrho_{YY}(t) &=& [\varrho_{YY}(0)+\varrho_{BB}(0)(1-e^{-\gamma t})]e^{-\gamma t}\\
\varrho_{GG}(t) &=& 1 - [\varrho_{XX}(0) +\varrho_{YY}(0) + \varrho_{BB}(0)(2-e^{-\gamma t})]e^{-\gamma t}
\end{eqnarray*}
The time $t=0$ corresponds to the time directly after each pulse.

From this, we can calculate the dynamics of all elements of the
density matrix, in other words, of all populations and coherences.
The FWM signal is theoretically extracted by analyzing the phase
dependence of the polarization. In general, all polarizations have
parts depending on different orders and combinations of the phases
$\varphi_i$ of the pulses. The two-pulse FWM for coherence dynamics
is given by the phase combination ($2\varphi_2-\varphi_1$), while
the three-pulse FWM for the population dynamics is characterized by
the phase combination ($\varphi_3+\varphi_2-\varphi_1$) which model
the heterodyning at ($2\Omega_2-\Omega_1$) and
($\Omega_3+\Omega_2-\Omega_1$). This identifies the polarization of
the FWM signal indicated by $p^{\rm FWM}$. For the sake of
simplicity, in the case of population dynamics we use
$\tau_{12}=0$~ps to mimic the short time delay between the first two
pulses. From the polarization the FWM signal $\mathcal S_{\nu
\nu^\prime}$ is obtained by a Fourier transform at the selected
frequency \be \mathcal S_{\nu \nu^\prime} = \left|
\left.\int_0^{\infty} p^{\rm FWM}\,e^{i\omega t}\,{\rm d}t\right.
\right|_{\omega=\omega_{\nu}-\omega_{\nu^\prime}}. \ee If the
polarization $\alpha$ is not along one axis of the QD, the signals
are added according to the angle of the heterodyning (reference)
beam $\alpha_r$ with $\mathcal S_{GXY} = \cos^2(\alpha_r) \mathcal
S_{GX} + \sin^2(\alpha_r) \mathcal S_{GY}$.

In the FWM signal, charge fluctuations can play an important role
leading to an inhomogeneous broadening via spectral wandering of
individual transitions. This phenomenon induces a photon echo in FWM
transients of single QDs, when probing the
coherence\,\cite{PattonPRB06, KasprzakNJP13, Mermillod16}. The
residual inhomogeneous broadening (i.e. up to several homogeneous
linewidths) can be included in the calculations by multiplying the
FWM-polarization with a Gaussian function \cite{KasprzakNJP13} as
follows: \be p^{\rm FWM} \to p^{\rm
FWM}\,e^{-\frac{(t-\tau_{12})^2}{2\sigma^2}} \ee For most cases the
inhomogeneous broadening can be neglected. We only included it to
model the data in Fig.~\ref{fig:cross} with $\sigma = 67$~ps, which
corresponds to an energetic broadening of $\hbar\sigma\approx10
$~\textmu eV.
\newpage


\section*{acknowledgements}
We acknowledge the support by the ERC Starting Grant PICSEN,
contract no.~306387. D.E.R. is grateful for financial support from the DAAD within the P.R.I.M.E. program.

\textbf{Author contributions:} Q.M., V.D. and J.K. performed
measurements. D.W. did the theoretical modeling. C.S, M.K. and S.H.
fabricated the sample. W.L. and G.N. provided technical assistance.
Q.M., D.W. D.E.R, T.K. and J.K. prepared the figures and wrote the
paper.


\begin{thebibliography}{42}%
\makeatletter
\providecommand \@ifxundefined [1]{%
 \@ifx{#1\undefined}
}%
\providecommand \@ifnum [1]{%
 \ifnum #1\expandafter \@firstoftwo
 \else \expandafter \@secondoftwo
 \fi
}%
\providecommand \@ifx [1]{%
 \ifx #1\expandafter \@firstoftwo
 \else \expandafter \@secondoftwo
 \fi
}%
\providecommand \natexlab [1]{#1}%
\providecommand \enquote  [1]{``#1''}%
\providecommand \bibnamefont  [1]{#1}%
\providecommand \bibfnamefont [1]{#1}%
\providecommand \citenamefont [1]{#1}%
\providecommand \href@noop [0]{\@secondoftwo}%
\providecommand \href [0]{\begingroup \@sanitize@url \@href}%
\providecommand \@href[1]{\@@startlink{#1}\@@href}%
\providecommand \@@href[1]{\endgroup#1\@@endlink}%
\providecommand \@sanitize@url [0]{\catcode `\\12\catcode
`\$12\catcode
  `\&12\catcode `\#12\catcode `\^12\catcode `\_12\catcode `\%12\relax}%
\providecommand \@@startlink[1]{}%
\providecommand \@@endlink[0]{}%
\providecommand \url  [0]{\begingroup\@sanitize@url \@url }%
\providecommand \@url [1]{\endgroup\@href {#1}{\urlprefix }}%
\providecommand \urlprefix  [0]{URL }%
\providecommand \Eprint [0]{\href }%
\providecommand \doibase [0]{http://dx.doi.org/}%
\providecommand \selectlanguage [0]{\@gobble}%
\providecommand \bibinfo  [0]{\@secondoftwo}%
\providecommand \bibfield  [0]{\@secondoftwo}%
\providecommand \translation [1]{[#1]}%
\providecommand \BibitemOpen [0]{}%
\providecommand \bibitemStop [0]{}%
\providecommand \bibitemNoStop [0]{.\EOS\space}%
\providecommand \EOS [0]{\spacefactor3000\relax}%
\providecommand \BibitemShut  [1]{\csname bibitem#1\endcsname}%
\let\auto@bib@innerbib\@empty
\bibitem [{\citenamefont {Bonadeo}\ \emph {et~al.}(1998)\citenamefont
  {Bonadeo}, \citenamefont {Erland}, \citenamefont {Gammon}, \citenamefont
  {Park}, \citenamefont {Katzer},\ and\ \citenamefont {Steel}}]{BonadeoSCI98}%
  \BibitemOpen
  \bibfield  {author} {\bibinfo {author} {\bibfnamefont {N.~H.}\ \bibnamefont
  {Bonadeo}}, \bibinfo {author} {\bibfnamefont {J.}~\bibnamefont {Erland}},
  \bibinfo {author} {\bibfnamefont {D.}~\bibnamefont {Gammon}}, \bibinfo
  {author} {\bibfnamefont {D.}~\bibnamefont {Park}}, \bibinfo {author}
  {\bibfnamefont {D.~S.}\ \bibnamefont {Katzer}}, \ and\ \bibinfo {author}
  {\bibfnamefont {D.~G.}\ \bibnamefont {Steel}},\ }\href@noop {} {\bibfield
  {journal} {\bibinfo  {journal} {Science}\ }\textbf {\bibinfo {volume}
  {282}},\ \bibinfo {pages} {1473} (\bibinfo {year} {1998})}\BibitemShut
  {NoStop}%
\bibitem [{\citenamefont {Michler}(2003)}]{Michler03}%
  \BibitemOpen
  \bibfield  {author} {\bibinfo {author} {\bibfnamefont {P.}~\bibnamefont
  {Michler}},\ }\href@noop {} {\emph {\bibinfo {title} {{Single quantum dots:
  Fundamentals, applications and new concepts}}}}\ (\bibinfo  {publisher}
  {Springer, Berlin},\ \bibinfo {year} {2003})\BibitemShut {NoStop}%
\bibitem [{\citenamefont {Ramsay}(2010)}]{RamsaySST10}%
  \BibitemOpen
  \bibfield  {author} {\bibinfo {author} {\bibfnamefont {A.~J.}\ \bibnamefont
  {Ramsay}},\ }\href@noop {} {\bibfield  {journal} {\bibinfo  {journal}
  {Semicond.\ Sci.~Technol.}\ }\textbf {\bibinfo {volume} {25}},\ \bibinfo
  {pages} {103001} (\bibinfo {year} {2010})}\BibitemShut {NoStop}%
\bibitem [{\citenamefont {Monniello}\ \emph {et~al.}(2013)\citenamefont
  {Monniello}, \citenamefont {Tonin}, \citenamefont {Hostein}, \citenamefont
  {Lemaitre}, \citenamefont {Martinez}, \citenamefont {Voliotis},\ and\
  \citenamefont {Grousson}}]{MonnielloPRL13}%
  \BibitemOpen
  \bibfield  {author} {\bibinfo {author} {\bibfnamefont {L.}~\bibnamefont
  {Monniello}}, \bibinfo {author} {\bibfnamefont {C.}~\bibnamefont {Tonin}},
  \bibinfo {author} {\bibfnamefont {R.}~\bibnamefont {Hostein}}, \bibinfo
  {author} {\bibfnamefont {A.}~\bibnamefont {Lemaitre}}, \bibinfo {author}
  {\bibfnamefont {A.}~\bibnamefont {Martinez}}, \bibinfo {author}
  {\bibfnamefont {V.}~\bibnamefont {Voliotis}}, \ and\ \bibinfo {author}
  {\bibfnamefont {R.}~\bibnamefont {Grousson}},\ }\href@noop {} {\bibfield
  {journal} {\bibinfo  {journal} {Phys.\ Rev.\ Lett.}\ }\textbf {\bibinfo
  {volume} {111}},\ \bibinfo {pages} {026403} (\bibinfo {year}
  {2013})}\BibitemShut {NoStop}%
\bibitem [{\citenamefont {Akimov}\ \emph {et~al.}(2006)\citenamefont {Akimov},
  \citenamefont {Andrews},\ and\ \citenamefont {Henneberger}}]{AkimovPRL06}%
  \BibitemOpen
  \bibfield  {author} {\bibinfo {author} {\bibfnamefont {I.~A.}\ \bibnamefont
  {Akimov}}, \bibinfo {author} {\bibfnamefont {J.~T.}\ \bibnamefont {Andrews}},
  \ and\ \bibinfo {author} {\bibfnamefont {F.}~\bibnamefont {Henneberger}},\
  }\href@noop {} {\bibfield  {journal} {\bibinfo  {journal} {Phys.\ Rev.\
  Lett.}\ }\textbf {\bibinfo {volume} {96}},\ \bibinfo {pages} {067401}
  (\bibinfo {year} {2006})}\BibitemShut {NoStop}%
\bibitem [{\citenamefont {Bacher}\ \emph {et~al.}(1999)\citenamefont {Bacher},
  \citenamefont {Weigand}, \citenamefont {Seufert}, \citenamefont
  {Kulakovskii}, \citenamefont {Gippius}, \citenamefont {Forchel},
  \citenamefont {Leonardi},\ and\ \citenamefont {Hommel}}]{BacherPRL99}%
  \BibitemOpen
  \bibfield  {author} {\bibinfo {author} {\bibfnamefont {G.}~\bibnamefont
  {Bacher}}, \bibinfo {author} {\bibfnamefont {R.}~\bibnamefont {Weigand}},
  \bibinfo {author} {\bibfnamefont {J.}~\bibnamefont {Seufert}}, \bibinfo
  {author} {\bibfnamefont {V.~D.}\ \bibnamefont {Kulakovskii}}, \bibinfo
  {author} {\bibfnamefont {N.~A.}\ \bibnamefont {Gippius}}, \bibinfo {author}
  {\bibfnamefont {A.}~\bibnamefont {Forchel}}, \bibinfo {author} {\bibfnamefont
  {K.}~\bibnamefont {Leonardi}}, \ and\ \bibinfo {author} {\bibfnamefont
  {D.}~\bibnamefont {Hommel}},\ }\href@noop {} {\bibfield  {journal} {\bibinfo
  {journal} {Phys.\ Rev.\ Lett.}\ }\textbf {\bibinfo {volume} {83}},\ \bibinfo
  {pages} {4417} (\bibinfo {year} {1999})}\BibitemShut {NoStop}%
\bibitem [{\citenamefont {Seguin}\ \emph {et~al.}(2005)\citenamefont {Seguin},
  \citenamefont {Schliwa}, \citenamefont {Rodt}, \citenamefont {P{\"o}tschke},
  \citenamefont {Pohl},\ and\ \citenamefont {Bimberg}}]{SeguinPRL05}%
  \BibitemOpen
  \bibfield  {author} {\bibinfo {author} {\bibfnamefont {R.}~\bibnamefont
  {Seguin}}, \bibinfo {author} {\bibfnamefont {A.}~\bibnamefont {Schliwa}},
  \bibinfo {author} {\bibfnamefont {S.}~\bibnamefont {Rodt}}, \bibinfo {author}
  {\bibfnamefont {K.}~\bibnamefont {P{\"o}tschke}}, \bibinfo {author}
  {\bibfnamefont {U.~W.}\ \bibnamefont {Pohl}}, \ and\ \bibinfo {author}
  {\bibfnamefont {D.}~\bibnamefont {Bimberg}},\ }\href@noop {} {\bibfield
  {journal} {\bibinfo  {journal} {Phys.\ Rev.\ Lett.}\ }\textbf {\bibinfo
  {volume} {95}},\ \bibinfo {pages} {257402} (\bibinfo {year}
  {2005})}\BibitemShut {NoStop}%
\bibitem [{\citenamefont {Young}\ \emph {et~al.}(2005)\citenamefont {Young},
  \citenamefont {Stevenson}, \citenamefont {Shields}, \citenamefont {Atkinson},
  \citenamefont {Cooper}, \citenamefont {Ritchie}, \citenamefont {Groom},
  \citenamefont {Tartakovskii},\ and\ \citenamefont {Skolnick}}]{YoungPRB05}%
  \BibitemOpen
  \bibfield  {author} {\bibinfo {author} {\bibfnamefont {R.~J.}\ \bibnamefont
  {Young}}, \bibinfo {author} {\bibfnamefont {R.~M.}\ \bibnamefont
  {Stevenson}}, \bibinfo {author} {\bibfnamefont {A.~J.}\ \bibnamefont
  {Shields}}, \bibinfo {author} {\bibfnamefont {P.}~\bibnamefont {Atkinson}},
  \bibinfo {author} {\bibfnamefont {K.}~\bibnamefont {Cooper}}, \bibinfo
  {author} {\bibfnamefont {D.~A.}\ \bibnamefont {Ritchie}}, \bibinfo {author}
  {\bibfnamefont {K.~M.}\ \bibnamefont {Groom}}, \bibinfo {author}
  {\bibfnamefont {A.~I.}\ \bibnamefont {Tartakovskii}}, \ and\ \bibinfo
  {author} {\bibfnamefont {M.~S.}\ \bibnamefont {Skolnick}},\ }\href@noop {}
  {\bibfield  {journal} {\bibinfo  {journal} {Phys.\ Rev.\ {\rm B}}\ }\textbf
  {\bibinfo {volume} {72}},\ \bibinfo {pages} {113305} (\bibinfo {year}
  {2005})}\BibitemShut {NoStop}%
\bibitem [{\citenamefont {Stevenson}\ \emph {et~al.}(2006)\citenamefont
  {Stevenson}, \citenamefont {Young}, \citenamefont {Atkinson}, \citenamefont
  {Cooper}, \citenamefont {Ritchie},\ and\ \citenamefont
  {Shields}}]{StevensonNAT06}%
  \BibitemOpen
  \bibfield  {author} {\bibinfo {author} {\bibfnamefont {R.~M.}\ \bibnamefont
  {Stevenson}}, \bibinfo {author} {\bibfnamefont {R.~J.}\ \bibnamefont
  {Young}}, \bibinfo {author} {\bibfnamefont {P.}~\bibnamefont {Atkinson}},
  \bibinfo {author} {\bibfnamefont {K.}~\bibnamefont {Cooper}}, \bibinfo
  {author} {\bibfnamefont {D.~A.}\ \bibnamefont {Ritchie}}, \ and\ \bibinfo
  {author} {\bibfnamefont {A.~J.}\ \bibnamefont {Shields}},\ }\href@noop {}
  {\bibfield  {journal} {\bibinfo  {journal} {Nature}\ }\textbf {\bibinfo
  {volume} {439}},\ \bibinfo {pages} {179} (\bibinfo {year}
  {2006})}\BibitemShut {NoStop}%
\bibitem [{\citenamefont {Akopian}\ \emph {et~al.}(2006)\citenamefont
  {Akopian}, \citenamefont {Lindner}, \citenamefont {Poem}, \citenamefont
  {Berlatzky}, \citenamefont {Avron}, \citenamefont {Gershoni}, \citenamefont
  {Gerardot},\ and\ \citenamefont {Petroff}}]{AkopianPRL06}%
  \BibitemOpen
  \bibfield  {author} {\bibinfo {author} {\bibfnamefont {N.}~\bibnamefont
  {Akopian}}, \bibinfo {author} {\bibfnamefont {N.~H.}\ \bibnamefont
  {Lindner}}, \bibinfo {author} {\bibfnamefont {E.}~\bibnamefont {Poem}},
  \bibinfo {author} {\bibfnamefont {Y.}~\bibnamefont {Berlatzky}}, \bibinfo
  {author} {\bibfnamefont {J.}~\bibnamefont {Avron}}, \bibinfo {author}
  {\bibfnamefont {D.}~\bibnamefont {Gershoni}}, \bibinfo {author}
  {\bibfnamefont {B.~D.}\ \bibnamefont {Gerardot}}, \ and\ \bibinfo {author}
  {\bibfnamefont {P.~M.}\ \bibnamefont {Petroff}},\ }\href@noop {} {\bibfield
  {journal} {\bibinfo  {journal} {Phys.\ Rev.\ Lett.}\ }\textbf {\bibinfo
  {volume} {96}},\ \bibinfo {pages} {130501} (\bibinfo {year}
  {2006})}\BibitemShut {NoStop}%
\bibitem [{\citenamefont {Ding}\ \emph {et~al.}(2010)\citenamefont {Ding},
  \citenamefont {Singh}, \citenamefont {Plumhof}, \citenamefont {Zander},
  \citenamefont {K{\v{r}}{\'a}pek}, \citenamefont {Chen}, \citenamefont
  {Benyoucef}, \citenamefont {Zwiller}, \citenamefont {D{\"o}rr}, \citenamefont
  {Bester}, \citenamefont {Bester}, \citenamefont {Rastelli},\ and\
  \citenamefont {Schmidt}}]{DingPRL10}%
  \BibitemOpen
  \bibfield  {author} {\bibinfo {author} {\bibfnamefont {F.}~\bibnamefont
  {Ding}}, \bibinfo {author} {\bibfnamefont {R.}~\bibnamefont {Singh}},
  \bibinfo {author} {\bibfnamefont {J.~D.}\ \bibnamefont {Plumhof}}, \bibinfo
  {author} {\bibfnamefont {T.}~\bibnamefont {Zander}}, \bibinfo {author}
  {\bibfnamefont {V.}~\bibnamefont {K{\v{r}}{\'a}pek}}, \bibinfo {author}
  {\bibfnamefont {Y.~H.}\ \bibnamefont {Chen}}, \bibinfo {author}
  {\bibfnamefont {M.}~\bibnamefont {Benyoucef}}, \bibinfo {author}
  {\bibfnamefont {V.}~\bibnamefont {Zwiller}}, \bibinfo {author} {\bibfnamefont
  {K.}~\bibnamefont {D{\"o}rr}}, \bibinfo {author} {\bibfnamefont
  {G.}~\bibnamefont {Bester}}, \bibinfo {author} {\bibfnamefont
  {G.}~\bibnamefont {Bester}}, \bibinfo {author} {\bibfnamefont
  {A.}~\bibnamefont {Rastelli}}, \ and\ \bibinfo {author} {\bibfnamefont
  {O.~G.}\ \bibnamefont {Schmidt}},\ }\href@noop {} {\bibfield  {journal}
  {\bibinfo  {journal} {Phys.\ Rev.\ Lett.}\ }\textbf {\bibinfo {volume}
  {104}},\ \bibinfo {pages} {067405} (\bibinfo {year} {2010})}\BibitemShut
  {NoStop}%
\bibitem [{\citenamefont {Tonin}\ \emph {et~al.}(2012)\citenamefont {Tonin},
  \citenamefont {Hostein}, \citenamefont {Voliotis}, \citenamefont {Grousson},
  \citenamefont {Lemaitre},\ and\ \citenamefont {Martinez}}]{ToninPRB12}%
  \BibitemOpen
  \bibfield  {author} {\bibinfo {author} {\bibfnamefont {C.}~\bibnamefont
  {Tonin}}, \bibinfo {author} {\bibfnamefont {R.}~\bibnamefont {Hostein}},
  \bibinfo {author} {\bibfnamefont {V.}~\bibnamefont {Voliotis}}, \bibinfo
  {author} {\bibfnamefont {R.}~\bibnamefont {Grousson}}, \bibinfo {author}
  {\bibfnamefont {A.}~\bibnamefont {Lemaitre}}, \ and\ \bibinfo {author}
  {\bibfnamefont {A.}~\bibnamefont {Martinez}},\ }\href@noop {} {\bibfield
  {journal} {\bibinfo  {journal} {Phys.\ Rev.\ {\rm B}}\ }\textbf {\bibinfo
  {volume} {85}},\ \bibinfo {pages} {155303} (\bibinfo {year}
  {2012})}\BibitemShut {NoStop}%
\bibitem [{\citenamefont {Langbein}\ and\ \citenamefont
  {Patton}(2006)}]{LangbeinOL06}%
  \BibitemOpen
  \bibfield  {author} {\bibinfo {author} {\bibfnamefont {W.}~\bibnamefont
  {Langbein}}\ and\ \bibinfo {author} {\bibfnamefont {B.}~\bibnamefont
  {Patton}},\ }\href@noop {} {\bibfield  {journal} {\bibinfo  {journal} {Opt.\
  Lett.}\ }\textbf {\bibinfo {volume} {31}},\ \bibinfo {pages} {1151} (\bibinfo
  {year} {2006})}\BibitemShut {NoStop}%
\bibitem [{\citenamefont {Voss}\ \emph {et~al.}(2006)\citenamefont {Voss},
  \citenamefont {R{\"u}ckmann}, \citenamefont {Gutowski}, \citenamefont {Axt},\
  and\ \citenamefont {Kuhn}}]{VossPRB06}%
  \BibitemOpen
  \bibfield  {author} {\bibinfo {author} {\bibfnamefont {T.}~\bibnamefont
  {Voss}}, \bibinfo {author} {\bibfnamefont {I.}~\bibnamefont {R{\"u}ckmann}},
  \bibinfo {author} {\bibfnamefont {J.}~\bibnamefont {Gutowski}}, \bibinfo
  {author} {\bibfnamefont {V.~M.}\ \bibnamefont {Axt}}, \ and\ \bibinfo
  {author} {\bibfnamefont {T.}~\bibnamefont {Kuhn}},\ }\href@noop {} {\bibfield
   {journal} {\bibinfo  {journal} {Phys.\ Rev.\ {\rm B}}\ }\textbf {\bibinfo
  {volume} {73}},\ \bibinfo {pages} {115311} (\bibinfo {year}
  {2006})}\BibitemShut {NoStop}%
\bibitem [{\citenamefont {Moody}\ \emph {et~al.}(2013)\citenamefont {Moody},
  \citenamefont {Singh}, \citenamefont {Li}, \citenamefont {Akimov},
  \citenamefont {Bayer}, \citenamefont {Reuter}, \citenamefont {Wieck},\ and\
  \citenamefont {Cundiff}}]{MoodyPRB13}%
  \BibitemOpen
  \bibfield  {author} {\bibinfo {author} {\bibfnamefont {G.}~\bibnamefont
  {Moody}}, \bibinfo {author} {\bibfnamefont {R.}~\bibnamefont {Singh}},
  \bibinfo {author} {\bibfnamefont {H.}~\bibnamefont {Li}}, \bibinfo {author}
  {\bibfnamefont {I.~A.}\ \bibnamefont {Akimov}}, \bibinfo {author}
  {\bibfnamefont {M.}~\bibnamefont {Bayer}}, \bibinfo {author} {\bibfnamefont
  {D.}~\bibnamefont {Reuter}}, \bibinfo {author} {\bibfnamefont {A.~D.}\
  \bibnamefont {Wieck}}, \ and\ \bibinfo {author} {\bibfnamefont {S.~T.}\
  \bibnamefont {Cundiff}},\ }\href@noop {} {\bibfield  {journal} {\bibinfo
  {journal} {Phys.\ Rev.\ {\rm B}}\ }\textbf {\bibinfo {volume} {87}},\
  \bibinfo {pages} {045313} (\bibinfo {year} {2013})}\BibitemShut {NoStop}%
\bibitem [{\citenamefont {Fras}\ \emph {et~al.}(2015)\citenamefont {Fras},
  \citenamefont {Mermillod}, \citenamefont {Nogues}, \citenamefont {Hoarau},
  \citenamefont {Schneider}, \citenamefont {Kamp}, \citenamefont {H{\"o}fling},
  \citenamefont {Langbein},\ and\ \citenamefont {Kasprzak}}]{Fras15}%
  \BibitemOpen
  \bibfield  {author} {\bibinfo {author} {\bibfnamefont {F.}~\bibnamefont
  {Fras}}, \bibinfo {author} {\bibfnamefont {Q.}~\bibnamefont {Mermillod}},
  \bibinfo {author} {\bibfnamefont {G.}~\bibnamefont {Nogues}}, \bibinfo
  {author} {\bibfnamefont {C.}~\bibnamefont {Hoarau}}, \bibinfo {author}
  {\bibfnamefont {C.}~\bibnamefont {Schneider}}, \bibinfo {author}
  {\bibfnamefont {M.}~\bibnamefont {Kamp}}, \bibinfo {author} {\bibfnamefont
  {S.}~\bibnamefont {H{\"o}fling}}, \bibinfo {author} {\bibfnamefont
  {W.}~\bibnamefont {Langbein}}, \ and\ \bibinfo {author} {\bibfnamefont
  {J.}~\bibnamefont {Kasprzak}},\ }\href@noop {} {\bibfield  {journal}
  {\bibinfo  {journal} {Nature Photon.}\ } \textbf {\bibinfo {volume} {10}},\
  \bibinfo {pages} {155-158} (\bibinfo {year}
  {2016})}\BibitemShut {NoStop}%
\bibitem [{\citenamefont {Mermillod}\ \emph {et~al.}(2016)\citenamefont
  {Mermillod}, \citenamefont {Jakubczyk}, \citenamefont {Delmonte},
  \citenamefont {Delga}, \citenamefont {Peinke}, \citenamefont {G\'{e}rard},
  \citenamefont {Claudon},\ and\ \citenamefont {Kasprzak}}]{Mermillod16}%
  \BibitemOpen
  \bibfield  {author} {\bibinfo {author} {\bibfnamefont {Q.}~\bibnamefont
  {Mermillod}}, \bibinfo {author} {\bibfnamefont {T.}~\bibnamefont
  {Jakubczyk}}, \bibinfo {author} {\bibfnamefont {V.}~\bibnamefont {Delmonte}},
  \bibinfo {author} {\bibfnamefont {A.}~\bibnamefont {Delga}}, \bibinfo
  {author} {\bibfnamefont {E.}~\bibnamefont {Peinke}}, \bibinfo {author}
  {\bibfnamefont {J.-M.}\ \bibnamefont {G\'{e}rard}}, \bibinfo {author}
  {\bibfnamefont {J.}~\bibnamefont {Claudon}}, \ and\ \bibinfo {author}
  {\bibfnamefont {J.}~\bibnamefont {Kasprzak}},\ }\href@noop {} {\bibfield
  {journal} {\bibinfo  {journal} {accepted in Phys. Rev. Lett.}\ } (\bibinfo {year}
  {2016})}\BibitemShut {NoStop}%
\bibitem [{\citenamefont {Gschrey}\ \emph {et~al.}(2015)\citenamefont
  {Gschrey}, \citenamefont {Thoma}, \citenamefont {Schnauber}, \citenamefont
  {Seifried}, \citenamefont {Schmidt}, \citenamefont {Wohlfeil}, \citenamefont
  {Kr{\"u}ger}, \citenamefont {Schulze}, \citenamefont {Heindel}, \citenamefont
  {Burger}, \citenamefont {Strittmatter}, \citenamefont {Rodt},\ and\
  \citenamefont {Reitzenstein}}]{GschreyNatComm15}%
  \BibitemOpen
  \bibfield  {author} {\bibinfo {author} {\bibfnamefont {M.}~\bibnamefont
  {Gschrey}}, \bibinfo {author} {\bibfnamefont {A.}~\bibnamefont {Thoma}},
  \bibinfo {author} {\bibfnamefont {P.}~\bibnamefont {Schnauber}}, \bibinfo
  {author} {\bibfnamefont {M.}~\bibnamefont {Seifried}}, \bibinfo {author}
  {\bibfnamefont {R.}~\bibnamefont {Schmidt}}, \bibinfo {author} {\bibfnamefont
  {B.}~\bibnamefont {Wohlfeil}}, \bibinfo {author} {\bibfnamefont
  {L.}~\bibnamefont {Kr{\"u}ger}}, \bibinfo {author} {\bibfnamefont {J.-H.}\
  \bibnamefont {Schulze}}, \bibinfo {author} {\bibfnamefont {T.}~\bibnamefont
  {Heindel}}, \bibinfo {author} {\bibfnamefont {S.}~\bibnamefont {Burger}},
  \bibinfo {author} {\bibfnamefont {A.}~\bibnamefont {Strittmatter}}, \bibinfo
  {author} {\bibfnamefont {S.}~\bibnamefont {Rodt}}, \ and\ \bibinfo {author}
  {\bibfnamefont {S.}~\bibnamefont {Reitzenstein}},\ }\href@noop {} {\bibfield
  {journal} {\bibinfo  {journal} {Nature Comm.}\ }\textbf {\bibinfo {volume}
  {6}} (\bibinfo {year} {2015})}\BibitemShut {NoStop}%
\bibitem [{\citenamefont {Vagov}\ \emph {et~al.}(2002)\citenamefont {Vagov},
  \citenamefont {Axt},\ and\ \citenamefont {Kuhn}}]{VagovPRB02}%
  \BibitemOpen
  \bibfield  {author} {\bibinfo {author} {\bibfnamefont {A.}~\bibnamefont
  {Vagov}}, \bibinfo {author} {\bibfnamefont {V.~M.}\ \bibnamefont {Axt}}, \
  and\ \bibinfo {author} {\bibfnamefont {T.}~\bibnamefont {Kuhn}},\ }\href@noop
  {} {\bibfield  {journal} {\bibinfo  {journal} {Phys.\ Rev.\ {\rm B}}\
  }\textbf {\bibinfo {volume} {66}},\ \bibinfo {pages} {165312} (\bibinfo
  {year} {2002})}\BibitemShut {NoStop}%
\bibitem [{\citenamefont {Axt}\ \emph {et~al.}(2005)\citenamefont {Axt},
  \citenamefont {Kuhn}, \citenamefont {Vagov},\ and\ \citenamefont
  {Peeters}}]{AxtPRB05}%
  \BibitemOpen
  \bibfield  {author} {\bibinfo {author} {\bibfnamefont {V.~M.}\ \bibnamefont
  {Axt}}, \bibinfo {author} {\bibfnamefont {T.}~\bibnamefont {Kuhn}}, \bibinfo
  {author} {\bibfnamefont {A.}~\bibnamefont {Vagov}}, \ and\ \bibinfo {author}
  {\bibfnamefont {F.~M.}\ \bibnamefont {Peeters}},\ }\href@noop {} {\bibfield
  {journal} {\bibinfo  {journal} {Phys.\ Rev.\ {\rm B}}\ }\textbf {\bibinfo
  {volume} {72}},\ \bibinfo {pages} {125309} (\bibinfo {year}
  {2005})}\BibitemShut {NoStop}%
\bibitem [{\citenamefont {Kasprzak}\ \emph {et~al.}(2011)\citenamefont
  {Kasprzak}, \citenamefont {Patton}, \citenamefont {Savona},\ and\
  \citenamefont {Langbein}}]{KasprzakNPho11}%
  \BibitemOpen
  \bibfield  {author} {\bibinfo {author} {\bibfnamefont {J.}~\bibnamefont
  {Kasprzak}}, \bibinfo {author} {\bibfnamefont {B.}~\bibnamefont {Patton}},
  \bibinfo {author} {\bibfnamefont {V.}~\bibnamefont {Savona}}, \ and\ \bibinfo
  {author} {\bibfnamefont {W.}~\bibnamefont {Langbein}},\ }\href@noop {}
  {\bibfield  {journal} {\bibinfo  {journal} {Nature Photon.}\ }\textbf
  {\bibinfo {volume} {5}},\ \bibinfo {pages} {57} (\bibinfo {year}
  {2011})}\BibitemShut {NoStop}%
\bibitem [{\citenamefont {Cundiff}(2012)}]{CundiffJOSAB12}%
  \BibitemOpen
  \bibfield  {author} {\bibinfo {author} {\bibfnamefont {S.~T.}\ \bibnamefont
  {Cundiff}},\ }\href@noop {} {\bibfield  {journal} {\bibinfo  {journal} {J.\
  Opt.\ Soc.\ Am.\ {\rm B}}\ }\textbf {\bibinfo {volume} {29}},\ \bibinfo
  {pages} {A69} (\bibinfo {year} {2012})}\BibitemShut {NoStop}%
\bibitem [{\citenamefont {Dai}\ \emph {et~al.}(2012)\citenamefont {Dai},
  \citenamefont {Richter}, \citenamefont {Li}, \citenamefont {Bristow},
  \citenamefont {Falvo}, \citenamefont {Mukamel},\ and\ \citenamefont
  {Cundiff}}]{DaiPRL12}%
  \BibitemOpen
  \bibfield  {author} {\bibinfo {author} {\bibfnamefont {X.}~\bibnamefont
  {Dai}}, \bibinfo {author} {\bibfnamefont {M.}~\bibnamefont {Richter}},
  \bibinfo {author} {\bibfnamefont {H.}~\bibnamefont {Li}}, \bibinfo {author}
  {\bibfnamefont {A.~D.}\ \bibnamefont {Bristow}}, \bibinfo {author}
  {\bibfnamefont {C.}~\bibnamefont {Falvo}}, \bibinfo {author} {\bibfnamefont
  {S.}~\bibnamefont {Mukamel}}, \ and\ \bibinfo {author} {\bibfnamefont
  {S.~T.}\ \bibnamefont {Cundiff}},\ }\href@noop {} {\bibfield  {journal}
  {\bibinfo  {journal} {Phys.\ Rev.\ Lett.}\ }\textbf {\bibinfo {volume}
  {108}},\ \bibinfo {pages} {193201} (\bibinfo {year} {2012})}\BibitemShut
  {NoStop}%
\bibitem [{\citenamefont {Langbein}(2010)}]{LangbeinRNC10}%
  \BibitemOpen
  \bibfield  {author} {\bibinfo {author} {\bibfnamefont {W.}~\bibnamefont
  {Langbein}},\ }\href@noop {} {\bibfield  {journal} {\bibinfo  {journal}
  {Rivista del nuovo cimento}\ }\textbf {\bibinfo {volume} {33}},\ \bibinfo
  {pages} {255} (\bibinfo {year} {2010})}\BibitemShut {NoStop}%
\bibitem [{\citenamefont {Patton}\ \emph {et~al.}()\citenamefont {Patton},
  \citenamefont {Woggon},\ and\ \citenamefont {Langbein}}]{PattonPRL05}%
  \BibitemOpen
  \bibfield  {author} {\bibinfo {author} {\bibfnamefont {B.}~\bibnamefont
  {Patton}}, \bibinfo {author} {\bibfnamefont {U.}~\bibnamefont {Woggon}}, \
  and\ \bibinfo {author} {\bibfnamefont {W.}~\bibnamefont {Langbein}},\
  }\href@noop {} {\bibfield  {journal} {\bibinfo  {journal} {Phys.\ Rev.\
  Lett.}\ }\textbf {\bibinfo {volume} {95}},\ \bibinfo
  {pages} {266401} (\bibinfo {year} {2005})}\BibitemShut {NoStop}%
\bibitem [{\citenamefont {Reiter}\ \emph {et~al.}(2014)\citenamefont {Reiter},
  \citenamefont {Kuhn}, \citenamefont {Gl{\"a}ssl},\ and\ \citenamefont
  {Axt}}]{ReiterJPC14}%
  \BibitemOpen
  \bibfield  {author} {\bibinfo {author} {\bibfnamefont {D.~E.}\ \bibnamefont
  {Reiter}}, \bibinfo {author} {\bibfnamefont {T.}~\bibnamefont {Kuhn}},
  \bibinfo {author} {\bibfnamefont {M.}~\bibnamefont {Gl{\"a}ssl}}, \ and\
  \bibinfo {author} {\bibfnamefont {V.~M.}\ \bibnamefont {Axt}},\ }\href@noop
  {} {\bibfield  {journal} {\bibinfo  {journal} {J.\ Phys.\ Condens.\ Matter}\
  }\textbf {\bibinfo {volume} {26}},\ \bibinfo {pages} {423203} (\bibinfo
  {year} {2014})}\BibitemShut {NoStop}%
\bibitem [{\citenamefont {Langbein}\ \emph {et~al.}(2004)\citenamefont
  {Langbein}, \citenamefont {Borri}, \citenamefont {Woggon}, \citenamefont
  {Stavarache}, \citenamefont {Reuter},\ and\ \citenamefont
  {Wieck}}]{LangbeinPRB04}%
  \BibitemOpen
  \bibfield  {author} {\bibinfo {author} {\bibfnamefont {W.}~\bibnamefont
  {Langbein}}, \bibinfo {author} {\bibfnamefont {P.}~\bibnamefont {Borri}},
  \bibinfo {author} {\bibfnamefont {U.}~\bibnamefont {Woggon}}, \bibinfo
  {author} {\bibfnamefont {V.}~\bibnamefont {Stavarache}}, \bibinfo {author}
  {\bibfnamefont {D.}~\bibnamefont {Reuter}}, \ and\ \bibinfo {author}
  {\bibfnamefont {A.~D.}\ \bibnamefont {Wieck}},\ }\href@noop {} {\bibfield
  {journal} {\bibinfo  {journal} {Phys.\ Rev.\ {\rm B}}\ }\textbf {\bibinfo
  {volume} {69}},\ \bibinfo {pages} {161301} (\bibinfo {year}
  {2004})}\BibitemShut {NoStop}%
\bibitem [{\citenamefont {Patton}\ \emph {et~al.}(2006)\citenamefont {Patton},
  \citenamefont {Langbein}, \citenamefont {Woggon}, \citenamefont {Maingault},\
  and\ \citenamefont {Mariette}}]{PattonPRB06}%
  \BibitemOpen
  \bibfield  {author} {\bibinfo {author} {\bibfnamefont {B.}~\bibnamefont
  {Patton}}, \bibinfo {author} {\bibfnamefont {W.}~\bibnamefont {Langbein}},
  \bibinfo {author} {\bibfnamefont {U.}~\bibnamefont {Woggon}}, \bibinfo
  {author} {\bibfnamefont {L.}~\bibnamefont {Maingault}}, \ and\ \bibinfo
  {author} {\bibfnamefont {H.}~\bibnamefont {Mariette}},\ }\href@noop {}
  {\bibfield  {journal} {\bibinfo  {journal} {Phys.\ Rev.\ {\rm B}}\ }\textbf
  {\bibinfo {volume} {73}},\ \bibinfo {pages} {235354} (\bibinfo {year}
  {2006})}\BibitemShut {NoStop}%
\bibitem [{\citenamefont {Kasprzak}\ and\ \citenamefont
  {Langbein}(2008)}]{KasprzakPRB08}%
  \BibitemOpen
  \bibfield  {author} {\bibinfo {author} {\bibfnamefont {J.}~\bibnamefont
  {Kasprzak}}\ and\ \bibinfo {author} {\bibfnamefont {W.}~\bibnamefont
  {Langbein}},\ }\href@noop {} {\bibfield  {journal} {\bibinfo  {journal}
  {Phys.\ Rev.\ {\rm B}}\ }\textbf {\bibinfo {volume} {78}},\ \bibinfo {pages}
  {041103} (\bibinfo {year} {2008})}\BibitemShut {NoStop}%
\bibitem [{\citenamefont {Kasprzak}\ \emph {et~al.}(2013)\citenamefont
  {Kasprzak}, \citenamefont {Portolan}, \citenamefont {Rastelli}, \citenamefont
  {Wang}, \citenamefont {Plumhof}, \citenamefont {Schmidt},\ and\ \citenamefont
  {Langbein}}]{KasprzakNJP13}%
  \BibitemOpen
  \bibfield  {author} {\bibinfo {author} {\bibfnamefont {J.}~\bibnamefont
  {Kasprzak}}, \bibinfo {author} {\bibfnamefont {S.}~\bibnamefont {Portolan}},
  \bibinfo {author} {\bibfnamefont {A.}~\bibnamefont {Rastelli}}, \bibinfo
  {author} {\bibfnamefont {L.}~\bibnamefont {Wang}}, \bibinfo {author}
  {\bibfnamefont {J.~D.}\ \bibnamefont {Plumhof}}, \bibinfo {author}
  {\bibfnamefont {O.~G.}\ \bibnamefont {Schmidt}}, \ and\ \bibinfo {author}
  {\bibfnamefont {W.}~\bibnamefont {Langbein}},\ }\href@noop {} {\bibfield
  {journal} {\bibinfo  {journal} {New J.\ Phys.}\ }\textbf {\bibinfo {volume}
  {15}},\ \bibinfo {pages} {055006} (\bibinfo {year} {2013})}\BibitemShut
  {NoStop}%
\bibitem [{\citenamefont {Kr{\"u}gel}\ \emph {et~al.}(2007)\citenamefont
  {Kr{\"u}gel}, \citenamefont {Vagov}, \citenamefont {Axt},\ and\ \citenamefont
  {Kuhn}}]{KrugelPRB07}%
  \BibitemOpen
  \bibfield  {author} {\bibinfo {author} {\bibfnamefont {A.}~\bibnamefont
  {Kr{\"u}gel}}, \bibinfo {author} {\bibfnamefont {A.}~\bibnamefont {Vagov}},
  \bibinfo {author} {\bibfnamefont {V.~M.}\ \bibnamefont {Axt}}, \ and\
  \bibinfo {author} {\bibfnamefont {T.}~\bibnamefont {Kuhn}},\ }\href@noop {}
  {\bibfield  {journal} {\bibinfo  {journal} {Phys.\ Rev.\ {\rm B}}\ }\textbf
  {\bibinfo {volume} {76}},\ \bibinfo {pages} {195302} (\bibinfo {year}
  {2007})}\BibitemShut {NoStop}%
\bibitem [{\citenamefont {Kuhlmann}\ \emph {et~al.}(2013)\citenamefont
  {Kuhlmann}, \citenamefont {Houel}, \citenamefont {Ludwig}, \citenamefont
  {Greuter}, \citenamefont {Reuter}, \citenamefont {Wieck}, \citenamefont
  {Poggio},\ and\ \citenamefont {Warburton}}]{KuhlmanNatPhys13}%
  \BibitemOpen
  \bibfield  {author} {\bibinfo {author} {\bibfnamefont {A.~V.}\ \bibnamefont
  {Kuhlmann}}, \bibinfo {author} {\bibfnamefont {J.}~\bibnamefont {Houel}},
  \bibinfo {author} {\bibfnamefont {A.}~\bibnamefont {Ludwig}}, \bibinfo
  {author} {\bibfnamefont {L.}~\bibnamefont {Greuter}}, \bibinfo {author}
  {\bibfnamefont {D.}~\bibnamefont {Reuter}}, \bibinfo {author} {\bibfnamefont
  {A.~D.}\ \bibnamefont {Wieck}}, \bibinfo {author} {\bibfnamefont
  {M.}~\bibnamefont {Poggio}}, \ and\ \bibinfo {author} {\bibfnamefont
  {R.}~\bibnamefont {Warburton}},\ }\href@noop {} {\bibfield  {journal}
  {\bibinfo  {journal} {Nature Phys.}\ }\textbf {\bibinfo {volume} {9}},\
  \bibinfo {pages} {570} (\bibinfo {year} {2013})}\BibitemShut {NoStop}%
\bibitem [{\citenamefont {Urbaszek}\ \emph {et~al.}(2003)\citenamefont
  {Urbaszek}, \citenamefont {Warburton}, \citenamefont {Karrai}, \citenamefont
  {Gerardot}, \citenamefont {Petroff},\ and\ \citenamefont
  {Garcia}}]{UrbaszekPRL03}%
  \BibitemOpen
  \bibfield  {author} {\bibinfo {author} {\bibfnamefont {B.}~\bibnamefont
  {Urbaszek}}, \bibinfo {author} {\bibfnamefont {R.~J.}\ \bibnamefont
  {Warburton}}, \bibinfo {author} {\bibfnamefont {K.}~\bibnamefont {Karrai}},
  \bibinfo {author} {\bibfnamefont {B.~D.}\ \bibnamefont {Gerardot}}, \bibinfo
  {author} {\bibfnamefont {P.~M.}\ \bibnamefont {Petroff}}, \ and\ \bibinfo
  {author} {\bibfnamefont {J.~M.}\ \bibnamefont {Garcia}},\ }\href@noop {}
  {\bibfield  {journal} {\bibinfo  {journal} {Phys.\ Rev.\ Lett.}\ }\textbf
  {\bibinfo {volume} {90}},\ \bibinfo {pages} {247403} (\bibinfo {year}
  {2003})}\BibitemShut {NoStop}%
\bibitem [{\citenamefont {Akimov}\ \emph {et~al.}(2005)\citenamefont {Akimov},
  \citenamefont {Kavokin}, \citenamefont {Hundt},\ and\ \citenamefont
  {Henneberger}}]{AkimovPRB05}%
  \BibitemOpen
  \bibfield  {author} {\bibinfo {author} {\bibfnamefont {I.~A.}\ \bibnamefont
  {Akimov}}, \bibinfo {author} {\bibfnamefont {K.~V.}\ \bibnamefont {Kavokin}},
  \bibinfo {author} {\bibfnamefont {A.}~\bibnamefont {Hundt}}, \ and\ \bibinfo
  {author} {\bibfnamefont {F.}~\bibnamefont {Henneberger}},\ }\href@noop {}
  {\bibfield  {journal} {\bibinfo  {journal} {Phys.\ Rev.\ {\rm B}}\ }\textbf
  {\bibinfo {volume} {71}},\ \bibinfo {pages} {075326} (\bibinfo {year}
  {2005})}\BibitemShut {NoStop}%
\bibitem [{\citenamefont {Kodriano}\ \emph {et~al.}(2014)\citenamefont
  {Kodriano}, \citenamefont {Schmidgall}, \citenamefont {Benny},\ and\
  \citenamefont {Gershoni}}]{KordianoSST14}%
  \BibitemOpen
  \bibfield  {author} {\bibinfo {author} {\bibfnamefont {Y.}~\bibnamefont
  {Kodriano}}, \bibinfo {author} {\bibfnamefont {E.~R.}\ \bibnamefont
  {Schmidgall}}, \bibinfo {author} {\bibfnamefont {Y.}~\bibnamefont {Benny}}, \
  and\ \bibinfo {author} {\bibfnamefont {D.}~\bibnamefont {Gershoni}},\
  }\href@noop {} {\bibfield  {journal} {\bibinfo  {journal} {Semicond.\
  Sci.~Technol.}\ }\textbf {\bibinfo {volume} {29}},\ \bibinfo {pages} {053001}
  (\bibinfo {year} {2014})}\BibitemShut {NoStop}%
\bibitem [{\citenamefont {Aharonovich}\ \emph {et~al.}(2011)\citenamefont
  {Aharonovich}, \citenamefont {Castelletto}, \citenamefont {Simpson},
  \citenamefont {Su}, \citenamefont {Greentree},\ and\ \citenamefont
  {Prawer}}]{AharonovichRPP11}%
  \BibitemOpen
  \bibfield  {author} {\bibinfo {author} {\bibfnamefont {I.}~\bibnamefont
  {Aharonovich}}, \bibinfo {author} {\bibfnamefont {S.}~\bibnamefont
  {Castelletto}}, \bibinfo {author} {\bibfnamefont {D.~A.}\ \bibnamefont
  {Simpson}}, \bibinfo {author} {\bibfnamefont {C.~H.}\ \bibnamefont {Su}},
  \bibinfo {author} {\bibfnamefont {A.~D.}\ \bibnamefont {Greentree}}, \ and\
  \bibinfo {author} {\bibfnamefont {S.}~\bibnamefont {Prawer}},\ }\href@noop {}
  {\bibfield  {journal} {\bibinfo  {journal} {Rep.\ Prog.\ Phys.}\ }\textbf
  {\bibinfo {volume} {74}},\ \bibinfo {pages} {076501} (\bibinfo {year}
  {2011})}\BibitemShut {NoStop}%
\bibitem [{\citenamefont {Tonndorf}\ \emph {et~al.}(2015)\citenamefont
  {Tonndorf}, \citenamefont {Schmidt}, \citenamefont {Schneider}, \citenamefont
  {Kern}, \citenamefont {Buscema}, \citenamefont {Steele}, \citenamefont
  {Castellanos-Gomez}, \citenamefont {van~der Zant}, \citenamefont
  {Michaelis~de Vasconcellos},\ and\ \citenamefont
  {Bratschitsch}}]{TonndorfOPTICA15}%
  \BibitemOpen
  \bibfield  {author} {\bibinfo {author} {\bibfnamefont {P.}~\bibnamefont
  {Tonndorf}}, \bibinfo {author} {\bibfnamefont {R.}~\bibnamefont {Schmidt}},
  \bibinfo {author} {\bibfnamefont {R.}~\bibnamefont {Schneider}}, \bibinfo
  {author} {\bibfnamefont {J.}~\bibnamefont {Kern}}, \bibinfo {author}
  {\bibfnamefont {M.}~\bibnamefont {Buscema}}, \bibinfo {author} {\bibfnamefont
  {G.~A.}\ \bibnamefont {Steele}}, \bibinfo {author} {\bibfnamefont
  {A.}~\bibnamefont {Castellanos-Gomez}}, \bibinfo {author} {\bibfnamefont
  {H.~S.~J.}\ \bibnamefont {van~der Zant}}, \bibinfo {author} {\bibfnamefont
  {S.}~\bibnamefont {Michaelis~de Vasconcellos}}, \ and\ \bibinfo {author}
  {\bibfnamefont {R.}~\bibnamefont {Bratschitsch}},\ }\href@noop {} {\bibfield
  {journal} {\bibinfo  {journal} {Optica}\ }\textbf {\bibinfo {volume} {2}},\
  \bibinfo {pages} {347} (\bibinfo {year} {2015})}\BibitemShut {NoStop}%
\bibitem [{\citenamefont {Koperski}\ \emph {et~al.}()\citenamefont {Koperski},
  \citenamefont {Nogajewski}, \citenamefont {Arora}, \citenamefont {Cherkez},
  \citenamefont {Mallet}, \citenamefont {Veuillen}, \citenamefont {Marcus},
  \citenamefont {Kossacki},\ and\ \citenamefont {Potemski}}]{KoperskiNN15}%
  \BibitemOpen
  \bibfield  {author} {\bibinfo {author} {\bibfnamefont {M.}~\bibnamefont
  {Koperski}}, \bibinfo {author} {\bibfnamefont {K.}~\bibnamefont
  {Nogajewski}}, \bibinfo {author} {\bibfnamefont {A.}~\bibnamefont {Arora}},
  \bibinfo {author} {\bibfnamefont {V.}~\bibnamefont {Cherkez}}, \bibinfo
  {author} {\bibfnamefont {P.}~\bibnamefont {Mallet}}, \bibinfo {author}
  {\bibfnamefont {J.-Y.}\ \bibnamefont {Veuillen}}, \bibinfo {author}
  {\bibfnamefont {J.}~\bibnamefont {Marcus}}, \bibinfo {author} {\bibfnamefont
  {P.}~\bibnamefont {Kossacki}}, \ and\ \bibinfo {author} {\bibfnamefont
  {M.}~\bibnamefont {Potemski}},\ }\href@noop {} {\bibfield  {journal}
  {\bibinfo  {journal} {Nature Nanotech.}\ }\textbf {\bibinfo {volume} {10}},\
  \bibinfo {pages} {503}}\BibitemShut {NoStop}%
\bibitem [{\citenamefont {He}\ \emph {et~al.}(2013{\natexlab{a}})\citenamefont
  {He}, \citenamefont {He}, \citenamefont {Wei}, \citenamefont {Wu},
  \citenamefont {Atat{\"u}re}, \citenamefont {Schneider}, \citenamefont
  {H{\"o}fling}, \citenamefont {Kamp}, \citenamefont {Lu},\ and\ \citenamefont
  {Pan}}]{HeNatNano13}%
  \BibitemOpen
  \bibfield  {author} {\bibinfo {author} {\bibfnamefont {Y.-M.}\ \bibnamefont
  {He}}, \bibinfo {author} {\bibfnamefont {Y.}~\bibnamefont {He}}, \bibinfo
  {author} {\bibfnamefont {Y.-J.}\ \bibnamefont {Wei}}, \bibinfo {author}
  {\bibfnamefont {D.}~\bibnamefont {Wu}}, \bibinfo {author} {\bibfnamefont
  {M.}~\bibnamefont {Atat{\"u}re}}, \bibinfo {author} {\bibfnamefont
  {C.}~\bibnamefont {Schneider}}, \bibinfo {author} {\bibfnamefont
  {S.}~\bibnamefont {H{\"o}fling}}, \bibinfo {author} {\bibfnamefont
  {M.}~\bibnamefont {Kamp}}, \bibinfo {author} {\bibfnamefont {C.-Y.}\
  \bibnamefont {Lu}}, \ and\ \bibinfo {author} {\bibfnamefont {J.-W.}\
  \bibnamefont {Pan}},\ }\href@noop {} {\bibfield  {journal} {\bibinfo
  {journal} {Nature Nanotech.}\ }\textbf {\bibinfo {volume} {8}},\ \bibinfo
  {pages} {213} (\bibinfo {year} {2013}{\natexlab{a}})}\BibitemShut {NoStop}%
\bibitem [{\citenamefont {He}\ \emph {et~al.}(2013{\natexlab{b}})\citenamefont
  {He}, \citenamefont {He}, \citenamefont {Wei}, \citenamefont {Jiang},
  \citenamefont {Chen}, \citenamefont {Xiong}, \citenamefont {Zhao},
  \citenamefont {Schneider}, \citenamefont {Kamp}, \citenamefont {H{\"o}fling},
  \citenamefont {Lu},\ and\ \citenamefont {Pan}}]{HePRL13}%
  \BibitemOpen
  \bibfield  {author} {\bibinfo {author} {\bibfnamefont {Y.}~\bibnamefont
  {He}}, \bibinfo {author} {\bibfnamefont {Y.-M.}\ \bibnamefont {He}}, \bibinfo
  {author} {\bibfnamefont {Y.-J.}\ \bibnamefont {Wei}}, \bibinfo {author}
  {\bibfnamefont {X.}~\bibnamefont {Jiang}}, \bibinfo {author} {\bibfnamefont
  {M.-C.}\ \bibnamefont {Chen}}, \bibinfo {author} {\bibfnamefont {F.-L.}\
  \bibnamefont {Xiong}}, \bibinfo {author} {\bibfnamefont {Y.}~\bibnamefont
  {Zhao}}, \bibinfo {author} {\bibfnamefont {C.}~\bibnamefont {Schneider}},
  \bibinfo {author} {\bibfnamefont {M.}~\bibnamefont {Kamp}}, \bibinfo {author}
  {\bibfnamefont {S.}~\bibnamefont {H{\"o}fling}}, \bibinfo {author}
  {\bibfnamefont {C.-Y.}\ \bibnamefont {Lu}}, \ and\ \bibinfo {author}
  {\bibfnamefont {J.-W.}\ \bibnamefont {Pan}},\ }\href@noop {} {\bibfield
  {journal} {\bibinfo  {journal} {Phys.\ Rev.\ Lett.}\ }\textbf {\bibinfo
  {volume} {111}},\ \bibinfo {pages} {237403} (\bibinfo {year}
  {2013}{\natexlab{b}})}\BibitemShut {NoStop}%
\bibitem [{\citenamefont {Maier}\ \emph {et~al.}(2014)\citenamefont {Maier},
  \citenamefont {Gold}, \citenamefont {Forchel}, \citenamefont {Gregersen},
  \citenamefont {M{\o}rk}, \citenamefont {H{\"o}fling}, \citenamefont
  {Schneider},\ and\ \citenamefont {Kamp}}]{MaierOptEx14}%
  \BibitemOpen
  \bibfield  {author} {\bibinfo {author} {\bibfnamefont {S.}~\bibnamefont
  {Maier}}, \bibinfo {author} {\bibfnamefont {P.}~\bibnamefont {Gold}},
  \bibinfo {author} {\bibfnamefont {A.}~\bibnamefont {Forchel}}, \bibinfo
  {author} {\bibfnamefont {N.}~\bibnamefont {Gregersen}}, \bibinfo {author}
  {\bibfnamefont {J.}~\bibnamefont {M{\o}rk}}, \bibinfo {author} {\bibfnamefont
  {S.}~\bibnamefont {H{\"o}fling}}, \bibinfo {author} {\bibfnamefont
  {C.}~\bibnamefont {Schneider}}, \ and\ \bibinfo {author} {\bibfnamefont
  {M.}~\bibnamefont {Kamp}},\ }\href@noop {} {\bibfield  {journal} {\bibinfo
  {journal} {Optics Express}\ }\textbf {\bibinfo {volume} {22}},\ \bibinfo
  {pages} {8136} (\bibinfo {year} {2014})}\BibitemShut {NoStop}%
\bibitem [{\citenamefont {Kasprzak}\ and\ \citenamefont
  {Langbein}(2009)}]{KasprzakPSSb09}%
  \BibitemOpen
  \bibfield  {author} {\bibinfo {author} {\bibfnamefont {J.}~\bibnamefont
  {Kasprzak}}\ and\ \bibinfo {author} {\bibfnamefont {W.}~\bibnamefont
  {Langbein}},\ }\href@noop {} {\bibfield  {journal} {\bibinfo  {journal}
  {Phys.\ Status Solidi {\rm B}}\ }\textbf {\bibinfo {volume} {246}},\ \bibinfo
  {pages} {820} (\bibinfo {year} {2009})}\BibitemShut {NoStop}%
\end{thebibliography}
\end{document}